\documentclass[aps,pra,twocolumn,superscriptaddress,showpacs,english,floats,eqsecnum,floatfix]{revtex4}

\usepackage{graphicx}
\usepackage{bm}
\usepackage{psfrag}
\usepackage{amsmath}
\usepackage{amsfonts}
\usepackage{amssymb}

\begin{document}

\title{Finite temperature correlations and density profiles of an \\
inhomogeneous interacting 1D Bose gas}
\author{K. V. Kheruntsyan}
\affiliation{ARC Centre of Excellence for Quantum-Atom Optics, Department of Physics,
University of Queensland, Brisbane, Qld 4072, Australia}
\author{D. M. Gangardt}
\affiliation{Laboratoire de Physique Th\'eorique et Mod\`eles Statistiques, Universit\'e
Paris Sud, 91405 Orsay Cedex, France}
\affiliation{Laboratoire Kastler-Brossel, Ecole Normale Sup\'erieure, 24 rue Lhomond,
75005 Paris, France}
\author{P. D. Drummond}
\affiliation{ARC Centre of Excellence for Quantum-Atom Optics, Department of Physics,
University of Queensland, Brisbane, Qld 4072, Australia}
\author{G. V. Shlyapnikov}
\affiliation{Laboratoire de Physique Th\'eorique et Mod\`eles Statistiques, Universit\'e
Paris Sud, 91405 Orsay Cedex, France}
\affiliation{Van der Waals - Zeeman Institute, University of Amsterdam, Valckenierstraat
65/67, 1018 XE Amsterdam, The Netherlands}
\date{\today}

\begin{abstract}
We calculate the density profiles and density correlation functions of the
one-dimensional Bose gas in a harmonic trap, using the exact
finite-temperature solutions for the uniform case, and applying a local
density approximation. The results are valid for a trapping potential which
is slowly varying relative to a correlation length. They allow a direct
experimental test of the transition from the weak coupling Gross-Pitaevskii
regime to the strong coupling, `fermionic' Tonks-Girardeau regime. We also
calculate the average two-particle correlation which characterizes the bulk
properties of the sample, and find that it can be well approximated by the
value of the local pair correlation in the trap center.
\end{abstract}

\maketitle





\section{Introduction}

The simplest investigations into a many-body system like a Bose-Einstein
condensate comprise studies of thermal equilibrium properties, and the
physics of small fluctuations around thermal equilibrium. For
one-dimensional systems, very similar behavior is found using either photons
in optical fibres or ultra-cold atoms in waveguides. Although techniques are
not yet as experimentally advanced in the latter case, preliminary theory
and some experimental measurements have already taken place. The atomic
systems have the advantage that relatively long interaction times, large
interaction strengths and low losses are possible, thus potentially allowing
stringent tests of underlying quantum correlations. In this paper, we extend
previous studies of correlations to include the experimentally realistic
case of atoms in a waveguide with a harmonic longitudinal confining
potential. The treatment is at finite temperature, and makes use of exact
results for the uniform gas, together with a local density approximation.

For strong radial confinement, these types of system are examples of
one-dimensional quantum ($1D$) gases \cite
{Goerlitz2001-Schreck2001-Greiner2001,Esslinger,Tolra-NIST-exp,
Weiss-exp,Bloch-and-Cazalilla}. They have the
important property that in many cases their energy eigenstates are exactly
solvable \cite
{Girardeau1960,LiebLiniger1963,Yang1969,Yang1970,Popov,Thacker1981,KorepinBook,Mattis}, 
resulting in a greatly increased fundamental understanding of the relevant
quantum field theory. For this reason, the study of 1D systems plays an
important role in the physics of quantum many-body systems. It is possible
to make first-principle predictions without introducing added approximations
like perturbation theory. This permits direct experimental tests of the
underlying many-body quantum physics, as has been demonstrated in photonics
with squeezed solitons in optical fibres \cite{soliton-squeezing}.

For ultra-cold atomic systems with repulsive interactions, the most
interesting and exciting feature is the predicted transition of an
interacting gas of bosons to a `fermionized' Tonks-Girardeau \cite
{Girardeau1960} regime at large coupling strengths and low densities.

We have recently made use of the known exact solutions to the uniform
one-dimensional (1D) interacting Bose gas problem to calculate the exact
local second-order correlation function at all densities and interaction
strengths \cite{Gangardt,KK-DG-PD-GS-2003,Deuar}. This is the most direct
indication of `fermionic' behaviour, since this correlation function is
strongly reduced at low density and strong coupling -- similar to the case
of fermions, where it vanishes exactly, due to the Pauli exclusion principle.

The first experimental evidence of reduced or `antibunched' correlations in
a $1D$ Bose gas has recently been demonstrated in Ref. \cite{Tolra-NIST-exp}
. However, current experiments typically take place in traps, with a
longitudinal trapping potential. Provided the trap potential varies slowly,
this environment is sufficiently close to a uniform one so that the exact
solutions can still be used locally, in an approximation called the local
density approximation.

In this paper, we make use of the local density approximation (LDA) to
calculate the density profile and finite temperature local pair correlation
function of a $1D$ Bose gas trapped in a harmonic potential. The results are
valid for sufficiently low longitudinal trap frequencies, and make use of
the exact solutions to the plane-wave Lieb-Liniger model \cite
{LiebLiniger1963} at finite temperature, together with the Hellmann-Feynman
theorem. We mostly focus on regimes with quantum degeneracy. This requires
temperatures $T\lesssim T_Q$, where $T_{Q}=N\hbar\omega_{z}$ is the
temperature of quantum degeneracy of the trapped sample as a whole, $N$ is
the total number of particles, and $\omega_{z}$ is the axial trap frequency.

Our main results show how `fermionization' can be readily detected through a
simple measurement of the pair correlation averaged over the trap. This is
very close to the correlation at the trap center as predicted \cite
{KK-DG-PD-GS-2003} using the Lieb-Liniger uniform model, and in principle
can be measured via photoassociation of trapped atoms, or other related
two-body inelastic processes whose rates are governed by the local pair
correlations \cite{Kagan1985}. In addition, an indirect measure of the pair
correlations can be obtained \cite{Gangardt,DG-GS-NewJournalPhysics} via the
measurement of three-body recombination rates as recently demonstrated
experimentally \cite{Tolra-NIST-exp}.

\section{One-dimensional Bose gas}

One dimensional quantum field theories have the important and useful
property that they are often exactly solvable. This is not generally the
case for higher-dimensional quantum field theories. Thus, the study of these
1D models can lead to an insight into the nature of quantum field theory for
interacting particles, that is not possible from the usual perturbative
approaches. In this section, we review the physics of these exact solutions
for interacting bosons in the uniform 1D case, and introduce the theoretical
framework for treating a non-uniform gas within the local density
approximation.

\subsection{Hamiltonian}

The study of exact solutions for the one-dimensional Bose gas started with
Girardeau's seminal work \cite{Girardeau1960} on hard-core, or impenetrable
bosons. In this model, there is a remarkable and exact correspondence
between the measurable correlation functions of free fermions, and those of
strongly interacting bosons. In the $1D$ Bose gas model with a
delta-function interaction, solved by Lieb and Liniger \cite{LiebLiniger1963}
, the particles can pass through each other, so they are no longer
impenetrable. This provides a realistic description of a wave-guide with
transverse dimensions larger than the `core' of a particle in the
wave-guide. Under these circumstances, there may only be a single relevant
transverse mode, yet particles are able to exchange their positions as they
propagate past each other.

Thus, we start by reviewing the theory of a gas of $N$ bosons interacting
via a delta-function potential in one dimension. The $1D$ Bose gas has a
short-range repulsive interaction between particles which is characterized
by just one coupling constant. In second quantization, the Hamiltonian is
\begin{align}
\hat{H} & =\frac{\hbar^{2}}{2m}\int dz\,\partial_{z}\hat{\Psi}^{\dagger
}\partial_{z}\hat{\Psi}+\frac{g}{2}\int dz\,\hat{\Psi}^{\dagger}\hat{\Psi }
^{\dagger}\hat{\Psi}\hat{\Psi}  \notag \\
& +\int dz\,V(z)\hat{\Psi}^{\dagger}\hat{\Psi},  \label{H}
\end{align}
where $\hat{\Psi}(z)$ is the bosonic field operator, $m$ is the atom mass, $
g>0$ is the coupling constant, and $V(z)$ is the trapping potential which we
assume is harmonic with $V(z)=m\omega_{z}^{2}z^{2}/2$, while $\omega_{z}$ is
the trap oscillation frequency in the axial direction. To treat the uniform
gas we set $V(z)=0$.

For Bose gases in highly elongated cylindrical traps ($\omega_{z}\ll
\omega_{\perp}$, where $\omega_{\perp}$ is the frequency of the transverse
harmonic potential) such that the sample can be described by the above $1D$
model, the coupling constant $g$ is expressed through the $3D$ scattering
length $a$ \cite{Olshanii98}. For a positive scattering length $a$ which is
much smaller than the amplitude of transverse ($x,y$-direction) zero point
oscillations, or the transverse harmonic oscillator length,
\begin{equation}
l_{\perp}=\sqrt{\hbar/m\omega_{\perp}},  \label{l-perp}
\end{equation}
one has
\begin{equation}
g\simeq\frac{2\hbar^{2}a}{ml_{\perp}^{2}} =2\hbar\omega_{\perp}a.  \label{g}
\end{equation}

The $1D$ regime is reached if $l_{\perp}$ is much smaller than the thermal
de Broglie wavelength $\Lambda_{T}=(2\pi\hbar^{2}/mT)^{1/2}$ and a
characteristic length scale $l_{c}$ \cite{lc} responsible for short-range
correlations. On the same grounds as at $T=0$ \cite{Gangardt}, one finds
that for fulfilling this requirement it is sufficient to satisfy the
inequalities
\begin{equation}
a\ll l_{\perp}\ll\{1/n(0),\,\Lambda_{T}\},  \label{1D-regime}
\end{equation}
where $n(0)=\left\langle \hat{\Psi}^{\dagger}(0)\hat{\Psi}(0)\right\rangle $
is the $1D$ (linear) density in the center of the trap, $z=0$.

\subsection{Ground-state solution for the uniform gas}

We now give a brief overview of the uniform ($V(z)=0$) Bose gas problem
describing a gas of $N$ bosons interacting via a pair-wise repulsive
delta-function potential in a $1D$ box of length $L$ with periodic boundary
condition \cite{LiebLiniger1963}. In the thermodynamic limit ($N$, $
L\rightarrow\infty$, while the $1D$ linear density $n=N/L$ is kept
constant), the solution to the energy eigenstates is found \cite
{LiebLiniger1963} using the Bethe ansatz \cite{Bethe}. In this solution, all
relative wave-functions are assumed to have a plane-wave form -- except for
finite changes in gradient at each collision where two particle coordinates
are equal:
\begin{align}
\left| \psi^{N}\right\rangle & =\int d^{N}\mathbf{z}\prod_{i=1}
^{N}e^{ik_{i}z_{i}}\left[ \prod_{j>i}\left( 1-\frac{ic}{k_{i}-k_{j} }
sgn(z_{i}-z_{j})\right) \right]  \notag \\
& \times\hat{\Psi}^{\dagger}(z_{1})\hat{\Psi}^{\dagger}(z_{2})...\hat{\Psi }
^{\dagger}(z_{N})\left| 0\right\rangle
\end{align}

Here, we have used units in which $\hbar=2m=1$, while introducing the
Lieb-Liniger notation \cite{LiebLiniger1963} of $c=mg/\hbar^2$. Also, $
sgn(z_{i}-z_{j})$ is the sign function, and $k_{i}$ is the
``quasi-momentum''. The quasi-momenta are determined from the delta-function
slope-change requirements, that is, $k_{j+1}-k_{j}$ is determined by the
boundary conditions at $z_{i}=z_{j}$. In the limit of a large sample, and
defining $k_{j+1}-k_{j}=1/[Lf(k_{j})]$, one can approximate $f(k_{j})$ by a
continuous function $f(k)$ which is the density of quasi-momenta. The
distribution of quasi-momenta is then obtained as the solution to the
following integral equation:
\begin{equation}
2\pi f(k)=1+\int_{-k_{F}}^{k_{F}}K(k-p)f(p)dp\,.  \label{LLequation}
\end{equation}
Here, the kernel function $K(k)$ is given by
\begin{equation}
K(k)=\frac{2c}{c^{2}+k^{2}},
\end{equation}
and $k_{F}$ is the maximum quasi-momentum which determines the particle
number density $n=N/L$ via
\begin{equation}
n=\int_{-k_{F}}^{k_{F}}f(k)dk
\end{equation}

The corresponding ground-state energy is given by
\begin{equation}
E_{0}=L\int_{-k_{F}}^{k_{F}}f(k)k^{2}dk,
\end{equation}
and is often written as $E_{0}=Nn^{2}e(\gamma)$, being an implicit function
of $n$, via a dimensionless function $e(\gamma)$ of the parameter $
\gamma=c/n $.

Restoring the physical units, this gives an energy per particle:
\begin{equation}
E_{0}/N=\frac{\hbar^{2}}{2m}n^{2}e(\gamma),  \label{E0}
\end{equation}
where
\begin{equation}
\gamma=\frac{mg}{\hbar^{2}n}.  \label{gamma}
\end{equation}

The dimensionless parameter $\gamma$ which characterizes the strength of
interactions is in fact the only parameter needed to describe the uniform $
1D $ Bose gas at zero temperature. The limit of $\gamma\ll1$ corresponds to
the weakly interacting Gross-Pitaevskii (GP) regime, where the mean-field
Bogoliubov theory works well. The opposite limit of $\gamma\gg 1$
corresponds to the strongly interacting or Tonks-Girardeau (TG) regime, and
as $\gamma\rightarrow\infty$ one regains Girardeau's results for
impenetrable bosons.

The solution to the ground state energy $E_{0}$ can be used, together with
the Hellmann-Feynman theorem \cite{Hellmann1933-Feynman1939}, for
calculating an important observable -- the normalized local pair correlation
\begin{equation}
g^{(2)}(0)=\frac{\left\langle \Psi^{\dagger}(z)\Psi^{\dagger}(z)\Psi
(z)\Psi(z)\right\rangle }{n^{2}}.  \label{g20}
\end{equation}

The pair correlation is found by taking the derivative of the ground state
energy with respect to the coupling constant $g$, owing to the fact that
\begin{equation}
\frac{dE_{0}}{dg}=\left\langle \frac{d\hat{H}}{dg}\right\rangle =\frac{L} {2}
\left\langle \Psi^{\dagger}(z)\Psi^{\dagger}(z)\Psi(z)\Psi(z)\right\rangle ,
\label{g2zeroT}
\end{equation}
so that
\begin{equation}
g^{(2)}(0)=\frac{de(\gamma)}{d\gamma}.  \label{g2derivative}
\end{equation}

The pair correlation $g^{(2)}(0)$ for the zero temperature uniform $1D$ Bose
gas has been calculated using the Lieb-Liniger exact solution \cite
{LiebLiniger1963} for $e(\gamma)$. The results are given in Ref. \cite
{Gangardt}. Here, we will extend these results (see Section III) to the case
of a trapped (non-uniform) Bose gas using the local density approximation,
and to finite temperatures as well.

\subsection{Uniform gas at finite temperature}

The excited states of the uniform 1D Bose gas can be calculated in a similar
way, with each excited state corresponding to the removal of a
quasi-momentum with $|k|<k_{F}$ -- called a hole -- and the creation of a
quasi-momentum with $|k|>k_{F}$. In 1969, C.N. Yang and C.P. Yang \cite
{Yang1969} worked out the finite-temperature density matrix solution for the
Lieb-Liniger model, by constructing the free energy and taking into account
the entropy of all the different excited states. This was used in a subsequent 
work \cite{Yang1970} to calculate numerically the pressure of the gas as a 
function of temperature.

At thermal equilibrium, we now assume that the density of quasi-momenta $f(k)$
has no upper cut-off, and that it consists of two types of terms -- occupied
quasi-momenta [with density $f_{p}(k)$] and unoccupied or `hole'
quasi-momenta [with density $f_{h}(k)$], i.e. $f(k)=f_{p}(k)+f_{h}(k)$. The
overall integral equation now has the form (using units in which $\hbar=2m=1$
and $c=mg/\hbar^2$)
\begin{align}
2\pi f(k) & =2\pi\left[ f_{p}(k)+f_{h}(k)\right]  \notag \\
& =1+\int_{-\infty}^{\infty}K(k-p)f_{p}(p)dp.
\end{align}

Hence, the particle density $n$ is obtained from the occupied or particle
quasi-momenta
\begin{equation}
n=\int_{-\infty}^{\infty}f_{p}(k)dk,  \label{nzT}
\end{equation}
while the total energy is now
\begin{equation}
E_0=L\int_{-\infty}^{\infty}f_{p}(k)k^{2}dk.
\end{equation}

However, there is also an entropy involved, since there are many
wavefunctions that are nearly the same, within a given range of values of $
f_{p}(k)$ and $f_{h}(k)$. In fact, the number of choices compatible with a
given $dk$ value is
\begin{equation}
\frac{\left[ f(k) Ldk\right] !}{\left[ f_{p}(k)Ldk\right] !\left[ f_{h}(k)Ldk
\right] !}.
\end{equation}

Thus, the entropy is
\begin{equation}
S =L\int_{-\infty}^{\infty}\left[ f \ln f -f_{p}\ln f_{p}-f_{h}\ln f_{h}
\right] dk.
\end{equation}

Minimizing the total free energy $F=E-TS$ gives the thermal equilibrium
distribution of holes and particles, where we choose temperature to be in
energy units, so that $k_B=1$. The minimization at a fixed average particle
number requires the use of a Lagrange multiplier $\mu$, and gives the result
that the distribution $f_{p}(k)$ satisfies the integral equation
\begin{equation}
2\pi f_{p}(k)\left[ 1+e^{\varepsilon(k)/T}\right] =1+\int_{-\infty}^{\infty
}K(k-p)f_{p}(p)dp,  \label{YY1}
\end{equation}
where the excitation spectrum $\varepsilon(k)$ is calculated from a second
integral equation
\begin{align}
\varepsilon(k) & =-\mu+k^{2}  \notag \\
& -\frac{T}{2\pi}\int_{-\infty}^{\infty}K(k-p)\ln\left( 1+e^{-\varepsilon
(p)/T}\right) dp\,.  \label{YY2}
\end{align}
Here, $\mu$ can be shown to coincide with the chemical potential of the
system, while the entropy and the free energy per particle are found from:
\begin{align}
S/N & =\frac{1}{n}\int_{-\infty}^{\infty}\left[ f(k) \ln\left(
1+e^{-\varepsilon(k)/T}\right) \right] dk  \notag \\
& +\frac{1}{nT}\int_{-\infty}^{\infty}f_{p}(k)\varepsilon(k)dk,
\label{entropy}
\end{align}
\begin{equation}
F/N=\mu-\frac{T}{2\pi n}\int_{-\infty}^{\infty}\ln\left( 1+e^{-\varepsilon
(k)/T}\right) dk.  \label{FoverN}
\end{equation}

In addition, using the thermodynamic identity $F=-PL+\mu N$, one can arrive
at the following simple result for the pressure of the gas
\begin{equation}
P(\mu,T)=\frac{T}{2\pi n}\int_{-\infty}^{\infty}\ln\left( 1+e^{-\varepsilon
(k)/T}\right) dk.  \label{P}
\end{equation}

To calculate the pair correlation $g^{(2)}(0)$ for a finite temperature gas
one can again use the Hellmann-Feynman theorem \cite{Hellmann1933-Feynman1939}
. Here, we consider the canonical partition function $Z=\exp (-F/T)=\mathrm{
Tr}\exp(-\hat{H}/T)$, where the trace is over the states of the system with
a fixed particle number $N$, at temperature $T$. Taking the derivative of $
F=-T\log Z$ with respect to the coupling constant $g$ we obtain
\begin{equation}
\frac{\partial F}{\partial g}=\frac{1}{Z}\mathrm{Tr}\left[ e^{-\hat{H} /T}
\frac{\partial\hat{H}}{\partial g}\right] =\frac{L}{2}\left\langle
\Psi^{\dagger}(z)\Psi^{\dagger}(z)\Psi(z)\Psi(z)\right\rangle .
\end{equation}
Introducing the free energy per particle $f=F/N$ and restoring the physical
units, this gives \cite{KK-DG-PD-GS-2003}:
\begin{equation}
g^{(2)}(0)=\frac{2}{Ln^{2}}\left( \frac{\partial F}{\partial g}\right)
_{N,T}=\frac{2m}{\hbar^{2}n}\left( \frac{\partial f(\gamma,\tau)} {
\partial\gamma}\right) _{n,\tau}.  \label{g2-canonical}
\end{equation}
Here, $\tau=T/T_{d}$ is a dimensionless temperature parameter, with $
T_{d}=\hbar^{2}n^{2}/2m$ being the temperature of quantum degeneracy for a
uniform gas. Hence, we have
\begin{equation}
\tau=\frac{2mT}{\hbar^{2}n^{2}}.  \label{tau1}
\end{equation}
The pair of the dimensionless parameters $\gamma$ and $\tau$ completely
characterize the properties of a finite temperature uniform gas.

Alternatively, the local pair correlation $g^{(2)}(0)$ can be calculated
within the grand canonical formalism. Here, we consider the grand canonical
partition function $\mathcal{Z}=\exp(-\Omega/T)=\mathrm{Tr}\exp[(\mu\hat
{N}
-\hat{H})/T]$, where $\Omega=F-\mu N=-PL$ is the grand canonical
thermodynamic potential and $P$ is the pressure. The trace is over the
states of the system, at a fixed chemical potential $\mu$ and temperature $T$
. Taking the derivative of $\Omega=-T\log\mathcal{Z}$ with respect to the
coupling $g$ we obtain:
\begin{align}
\frac{\partial\Omega}{\partial g} & =-T\frac{\partial\log\mathcal{Z} }{
\partial g}  \notag \\
& =-\frac{1}{\mathcal{Z}}{\text{Tr}}\left\{ \frac{\partial(\mu\hat{N} -\hat{H
})}{\partial g}\exp[(\mu\hat{N}-\hat{H})/T]\right\}  \notag \\
& =\frac{L}{2}\left\langle \Psi^{\dagger}(z)\Psi^{\dagger}(z)\Psi
(z)\Psi(z)\right\rangle \,.
\end{align}

Thus, the normalized pair correlation $g^{(2)}(0)$ can be calculated using
\begin{equation}
g^{(2)}(0)=\frac{2}{Ln^{2}}\left( \frac{\partial\Omega}{\partial g}\right)
_{\mu,T}=-\frac{1}{n^{2}}\left( \frac{\partial P}{\partial g}\right)
_{\mu,T}\;.  \label{g2-grand}
\end{equation}
This requires the use of Eq. (\ref{P}) for the pressure, which in turn is
found after solving the Yang-Yang integral equations (\ref{YY1}) and (\ref
{YY2}).

The local pair correlation for a finite temperature uniform gas has been
first calculated in Ref. \cite{KK-DG-PD-GS-2003} using the exact solutions
to the Yang-Yang integral equations (\ref{YY1}) and (\ref{YY2}), together
with Eqs. (\ref{FoverN}) and (\ref{g2-canonical}). In Sections IV-VII, we
will extend these results to the case of a trapped gas using the local
density approximation.

\subsection{Quasi-uniform approximation}

In a quasi-uniform approximation, we suppose that the system can be divided
into small regions of size $\Delta z$ which is larger than a characteristic
short-range correlation length $l_{c}$. In each of this regions we assume
that the inhomogeneity of the gas is negligible so that it can be treated as
a uniform gas.

In this case, the trapping potential $V(z)$ is replaced by a step-like
function $V(z_{j})=m\omega_{z}^{2}z_{j}^{2}/2$ which is constant within each
region from $z_{j}$ to $z_{j+1}$ and undergoes step-like changes at the
boundaries of the adjacent regions. Here, the size $\Delta z$ takes the role
of the length $L$ from the Yang-Yang solution that applies to each region.

We now consider an ansatz in which the overall density matrix has the
structure of an outer-product of canonical solutions, with $N_{j}$ being the
average number of particles in the $j$-th region:
\begin{equation}
\widehat{\rho}^{N}=\widehat{\rho}^{N_{1}}(z_{1})\widehat{\rho}^{N_{2}}
(z_{2})...\widehat{\rho}^{N_{n}}(z_{n}).
\end{equation}

Next we look for an approximate solution in which the effective Hamiltonian
is assumed to introduce no coupling between the regions. To obtain this we
must now minimize the total free energy given by
\begin{equation}
F_{N}=\sum_{j=1}^{n}\left( E_{j}-TS_{j}\right) .
\end{equation}
This requires us to include a constraint on the total particle number:
\begin{equation}
N=\sum_{j=1}^{n}N_{j}.
\end{equation}
Hence, it is appropriate to use a Lagrangian formulation with
\begin{equation}
\mathcal{L}=\sum_{j=1}^{n}\left( E_{j}-TS_{j}-\mu_{0}N_{j}\right).
\end{equation}

We note here that the Lagrangian $\mathcal{L}$ is now simply a sum over
independent regions, with each term corresponding to that for a single
uniform Bose gas. As we are only constraining the total particle number, not
the number in each region, the chemical potential is the same for each term.
Since there is no explicit coupling between the regions, the Lagrangian is
minimized when we satisfy the Yang-Yang equations in each separate region,
but with the same (global) chemical potential $\mu_{0}$ at all locations.

\subsection{Local density approximation (LDA)}

In more detail, we have shown that for a large system, where the density
profile varies in a smooth way, the system behaves locally as a piece of a
uniform gas. This can be described locally as a uniform gas with chemical
potential equal to the local effective chemical potential
\begin{equation}
\mu(z)=\mu_{0}-V(z)=\mu_{0}-\frac{1}{2}m\omega_{z}^{2}z^{2},  \label{mu-z}
\end{equation}
where $\mu_{0}$ is the global equilibrium chemical potential.

For the LDA to be valid, the short-range correlation length $l_{c}(z)$
should be much smaller than the characteristic inhomogeneity length $
l_{inh}(z)$. These length scales depend on the displacement from the trap
center $z$, and the LDA validity criterion reads:
\begin{equation}
l_{c}(z)\ll l_{inh}(z)\equiv\frac{n(z)}{|dn(z)/dz|}.  \label{LDA-criterion}
\end{equation}

The short-range correlation length $l_{c}(z)$ is defined locally via the
density distribution $n(z)$. At low temperatures, the correlation length $
l_{c}(z)$ can in general (irrespective of the interaction strength) be
expressed via the local chemical potential $\mu(z)$:
\begin{equation}
l_{c}(z)=\frac{\hbar}{\sqrt{m\mu(z)}},~~~~~~(T\ll T_{Q}).  \label{l-c-mu}
\end{equation}
In the weakly interacting Gross-Pitaevskii (GP) regime the relation between
the chemical potential $\mu(z)$ and the density $n(z)$ is $\mu(z)=gn(z)$,
and we obtain that the correlation length coincides with the healing length $
l_{c}(z)=\hbar/\sqrt{mgn(z)}$. In the strongly interacting Tonks-Girardeau
(TG) regime one has $\mu(z)=\pi^{2}\hbar^{2}n^{2}(z)/(2m)$, so that $
l_{c}(z)\sim 1/n(z)$, neglecting the numerical factor of order $1$. At high
temperatures, $l_{c}(z)$ is of the order of the thermal de Broglie
wavelength $\Lambda_{T}$.

The condition (\ref{LDA-criterion}) is sufficient for using the LDA for
calculating the density profiles and local correlation functions. The reason
is that these correlations, in particular the two-particle correlation $
g^{(2)}(0)$, are determined by the contribution of excitations which have
energies of the order of the chemical potential and wavelength of the order
of $l_{c}$. However, this is not the case for all correlation functions. For
example, calculation of the finite-temperature single-particle correlation
function would require a strong LDA condition in which
the sample size was much larger than the phase correlation
distance \cite{Petrov}. In this sense, one may call Eq. (\ref{LDA-criterion}) the
`weak' LDA criterion. However, within the LDA, no correlation function can 
be calculated reliably over distance scales that are comparable 
to the sample size.

Thus, the `weak' criterion of validity of the LDA requires that variations
of the density occur on a length scale that is much larger than $l_{c}(z)$,
in which case the gas is treated locally as a piece of a uniform gas. From
the definition of $l_{inh}(z)$, one can easily see that the LDA is easier to
satisfy in the center of the trap where the density profile is almost flat
than near the tails of the distribution where the density drops rapidly.
However, for measurements that average over an entire trap, it is the
central region that plays the most important role.

\section{Zero-temperature trapped gas}

\subsection{LDA criterion at $T=0$}

Here, we analyze the implications of the LDA criterion (\ref{LDA-criterion})
for a zero temperature gas. At $T=0$, a uniform $1D$ Bose gas can be
characterized by a single dimensionless interaction parameter $\gamma$, Eq. (
\ref{gamma}). Depending on its value, one has two well-known and physically
distinct regimes of quantum degeneracy. For $\gamma\ll1$, i.e. at weak
couplings or high densities, the gas is in a coherent or Gross-Pitaevskii
(GP) regime. In this regime, long-range order is destroyed by
long-wavelength phase fluctuations \cite{Haldane,Mermin1966,Hohenberg,Petrov}
and the equilibrium state is a quasi-condensate characterized by suppressed
density fluctuations. For strong couplings or low densities, $\gamma\gg1$,
the gas reaches the strongly interacting or Tonks-Girardeau (TG) regime and
undergoes ``fermionization'' \cite{Girardeau1960,LiebLiniger1963}. The term
``fermionization'' is used here in the sense that the wave function strongly
decreases as particles approach each other.

For a trapped (non-uniform) gas one can introduce a local interaction
parameter
\begin{equation}
\gamma(z)=\frac{mg}{\hbar^{2}n(z)},  \label{gamma-z}
\end{equation}
which changes with the density distribution $n(z)$ and can be used for
characterizing the local properties of the gas.

From the definition of $\gamma(z)$ it is clear that as one moves from the
center of the trap towards the tails of the density distribution where $
n(z)\rightarrow0$, the gas either enters the TG regime where $\gamma(z)\gg1$, 
or else the LDA itself breaks down.

Moreover, in the $T=0$ case the Lieb-Liniger solution within the LDA gives a
density profile that vanishes [$n(z)=0$] beyond a certain distance $R$ from
the origin \cite{Olshanii2001,Kolomeisky2000}. This distance is called the
Thomas-Fermi radius and it is determined from the condition $
\mu(R)=\mu_{0}-m\omega_{z}^{2}R^{2}/2=0$ which gives:
\begin{equation}
R=\left( \frac{2\mu_{0}}{m\omega_{z}^{2}}\right) ^{1/2}.  \label{R-TF-gen}
\end{equation}

Since $n(z)$ vanishes exactly at $|z|\geq R$ \cite{comment-LDA}, it is clear
that the LDA criterion (\ref{LDA-criterion}) can only be satisfied up to a
certain maximum distance from the trap center $|z|\simeq R-\delta z$,
displaced from $R$ by $\delta z$ ($\delta z\ll R$). We would like therefore
to determine the displacement $\delta z$ such that the LDA is valid for $
0\leq|z|\leq R-\delta z$ and breaks down beyond $|z|\simeq R-\delta z$.

As we are interested in calculating the density profiles and the local
two-particle correlation function, we will only focus on the `weak' LDA
criterion, Eq. (\ref{LDA-criterion}). First, we rewrite the inhomogeneity
length scale from Eq. (\ref{LDA-criterion}) in the following equivalent form
\begin{equation}
l_{inh}(z)=n(z)\left| \frac{d\mu(z)/dn(z)}{d\mu(z)/dz}\right| .
\label{l-inh-alt}
\end{equation}

Using the explicit expression $\mu(z)=\mu_{0}-m\omega_{z}^{2}z^{2}/2$ and
taking its derivative, we obtain at $|z|\simeq R-\delta z$ ($\delta z\ll R$):
\begin{equation}
l_{inh}(z)\simeq\frac{n(z)}{m\omega_{z}^{2}R}\left| \frac{d\mu(z)} {dn(z)}
\right| .  \label{l-inh-approx}
\end{equation}

Combining Eqs. (\ref{l-c-mu}) and (\ref{l-inh-approx}), one can rewrite the
LDA criterion $l_{c}(z)\ll l_{inh}(z)$ in the following equivalent form
(again neglecting numerical factors of order one):
\begin{equation}
\left( \frac{\mu(z)}{\mu_{0}}\right) ^{3/2}\frac{\mu_{0}}{\hbar\omega} \frac{
d\ln\mu(z)}{d\ln n(z)}\gg1.
\end{equation}

Next, we note that in the limiting GP and TG regimes the derivative $d\ln
\mu(z)/d\ln n(z)$ is equal, respectively, to one and two, so in general $
1\leq d\ln\mu(z)/d\ln n(z)\leq2$. Therefore, the exact numerical value of
this quantity can be replaced by unity in all regimes, and the LDA criterion
becomes
\begin{equation}
\left( \frac{\mu(z)}{\mu_{0}}\right) ^{3/2}\frac{\mu_{0}}{\hbar\omega_{z} }
\gg1.
\end{equation}

Finally, expanding $\mu(z)$ near the edge of the cloud where $|z|\simeq
R-\delta z$ ($\delta z\ll R$), we obtain that $\mu(z)/\mu_{0}\simeq\delta
z/R $, so that the criterion of applicability of the LDA is reduced to a
simple requirement:
\begin{equation}
\frac{\delta z}{R}\gg\left( \frac{\hbar\omega_{z}}{\mu_{0}}\right) ^{2/3}.
\label{LDA-criterion-final}
\end{equation}

For $\gamma(0)\ll1$, the local value of $\gamma(z)$ at $z\simeq R-\delta z$
is
\begin{equation}
\gamma(z)\simeq\frac{mgR}{\hbar^{2}n(0)\delta z}\simeq\gamma(0)\left( \frac{
\mu_{0}}{\hbar\omega_{z}}\right) ^{2/3}.
\end{equation}
We see that if $\gamma(0)\ll\left( \hbar\omega_{z}/\mu_{0}\right) ^{2/3}$
then $\gamma(z)\ll1$ at $z\simeq R-\delta z$, implying that the gas stays in
the GP regime at all locations $z$ until the LDA breaks down. If, on the
other hand, $\gamma(0)\gg\left( \hbar\omega_{z}/\mu_{0}\right) ^{2/3}$ then $
\gamma(z)\gg1$ at $z\simeq R-\delta z$ so that the gas first approaches the
TG regime and then the LDA breaks down.

For $\gamma(0)\gg1$ the gas is in the TG regime at all locations $z$, until
the LDA breaks down. This is because $\gamma(z)$ is always larger than $
\gamma(0)$, and hence $\gamma(z)\gg1$.

In the limiting cases of $\gamma(0)\ll1$ and $\gamma(0)\gg1$, the LDA
criterion (\ref{LDA-criterion-final}) can be conveniently rewritten in terms
of $\gamma(0)$ and the total number of particles $N$. In doing so we use the
fact that the chemical potential is given by $\mu_{0}=gn(0)$ for $\gamma
(0)\ll1$, and by $\mu_{0}=\pi^{2}\hbar^{2}n^{2}(0)/(2m)$ for $\gamma(0)\gg1$
. In addition, we use the known relationship between the peak density $n(0)$
and $N$ in each case [see Eqs. (\ref{TF-GP-density}) and (\ref{TF-TG-density}
) below]. As a result, we obtain that the LDA criterion (\ref
{LDA-criterion-final}) can be rewritten as follows, in the GP and TG
regimes, respectively:
\begin{align}
\frac{\delta z}{R} & \gg\left( \frac{\hbar\omega_{z}}{gn(0)}\right)
^{2/3}\simeq\frac{1}{\gamma(0)^{1/3}N^{2/3}},\;[\gamma(0)\ll1],
\label{LDA-GP} \\
\frac{\delta z}{R} & \gg\left( \frac{2m\omega_{z}}{\pi^{2}\hbar^{2} n^{2}(0)}
\right) ^{2/3}=\frac{1}{N^{2/3}},\;\;\;\;\;[\gamma(0)\gg1].  \label{LDA-TG}
\end{align}

As we see, for any small but finite $\gamma(0)$ in the GP regime, the right
hand side of Eq. (\ref{LDA-GP}) can be made small by increasing the total
number of particles $N$. For a fixed coupling $g$ and a constant density $
n(0)$ [such that $\gamma(0)$ stays constant], the increase of the particle
number $N$ has to be accompanied by a reduction of the trap frequency $
\omega_{z}$. Thus, the ratio $\delta z/R$ can also be made small, so that
that the LDA criterion in the GP regime [$\gamma(0)\ll1$] is satisfied for
almost the entire sample, up to the location $z=R-\delta z$ very close to
the edge of the cloud. Similar considerations apply to the TG regime [$
\gamma(0)\gg1$], where the requirement on the (large) total number of
particles $N$ is less stringent than in the GP regime.

\subsection{Pair correlations at $T=0$}

Here we discuss the local pair correlation
\begin{equation}
g^{(2)}(z,z)\equiv\frac{\left\langle \hat{\Psi}^{\dagger}(z)\hat{\Psi }
^{\dagger}(z)\hat{\Psi}(z)\hat{\Psi}(z)\right\rangle }{n^{2}(z)}  \label{g2z}
\end{equation}
in a zero temperature trapped gas within the LDA. The calculations are done
using the solution to the Lieb-Liniger equation (\ref{LLequation}) and the
Hellmann-Feynman theorem \cite{Hellmann1933-Feynman1939}, Eq. (\ref
{g2derivative}). Here, $g^{(2)}(0)$ is now replaced by $g^{(2)}(z,z)$ and $
\gamma$ is to be understood as the local value of $\gamma(z)$. Thus, to
calculate $g^{(2)}(z,z)$ as a function of the distance from the trap centre,
one can use the uniform results in which the interaction parameter $
\gamma(z) $ is found from the density profile $n(z)$, for different values
of $\gamma(0)$.

The implementation of the LDA, using the local effective chemical potential $
\mu(z)$, Eq. (\ref{mu-z}), is carried out by means of first calculating the
chemical potential $\mu$ as a function of $n$, and then inverting this
dependence for obtaining $n(z)$. This gives $n(z)$ [and hence $\gamma(z)$]
as a function of $\mu(z)$, for a given value of the interaction parameter $
\gamma(0)$ at the trap center.

Depending on the value of the coordinate-dependent interaction parameter $
\gamma(z)$, Eq.~(\ref{gamma-z}), we have the following limiting behavior of
the pair correlation function.

In the Gross-Pitaevskii (GP) limit of a weakly interacting gas, the pair
correlation in the uniform case is $g^{(2)}\simeq 1-2\sqrt{\gamma }/\pi $, $
\gamma \ll 1$ \cite{Gangardt}. For a trapped gas, replacing $\gamma $ by $
\gamma (z)$ gives:
\begin{equation}
g^{(2)}(z,z)\simeq 1-\frac{2}{\pi }\sqrt{\frac{\gamma (0)}{1-z^{2}/R^{2}}}
,\;\;\;\gamma (z)\ll 1.  \label{g2GP}
\end{equation}
where we have used the relationship between $\gamma (z)$ and $n(z)$ and the
fact that in the GP regime the density profile is given by the familiar
Thomas-Fermi parabola:
\begin{equation}
n(z)=n(0)(1-z^{2}/R^{2}),  \label{TF-parabola}
\end{equation}
and $n(z)=0$ for $|z|\geq R$. Here, the peak density $n(0)$ and the radius $R
$ are given by:
\begin{align}
n(0)& =\left( \frac{9mN^{2}\omega _{z}^{2}}{32g}\right) ^{1/3},
\label{TF-GP-density} \\
R& =\left( \frac{3Ng}{2m\omega _{z}^{2}}\right) ^{1/3}.  \label{TF-GP-radius}
\end{align}

In the Tonks-Girardeau (TG) limit of strong interactions, the uniform gas
pair correlation is $g^{(2)}\simeq 4\pi ^{2}/(3\gamma ^{2})$, $\gamma \gg 1$
\cite{Gangardt}. In the trapped gas case, replacing $\gamma $ by $\gamma (z)$
gives
\begin{equation}
g^{(2)}(z,z)\simeq \frac{4\pi ^{2}(1-z^{2}/R^{2})}{3\gamma ^{2}(0)}
,\;\;\;\gamma (z)\gg 1,  \label{g2TG}
\end{equation}
where we again used the relationship between $\gamma (z)$ and $n(z)$ and the
fact that the density profile in the TG regime is given by the square root
of parabola:
\begin{equation}
n(z)=n(0)(1-z^{2}/R^{2})^{1/2},  \label{TF-parabola-TG}
\end{equation}
and $n(z)=0$ for $|z|\geq R$. Here, the peak density $n(0)$ and the radius $
R $ are
\begin{align}
n(0) & =\left( \frac{2mN\omega_{z}}{\pi^{2}\hbar}\right) ^{1/2} ,
\label{TF-TG-density} \\
R & =\left( \frac{2\hbar N}{m\omega_{z}}\right) ^{1/2}.  \label{TF-TG-radius}
\end{align}

\begin{figure}[ptb]
\includegraphics[width=6cm]{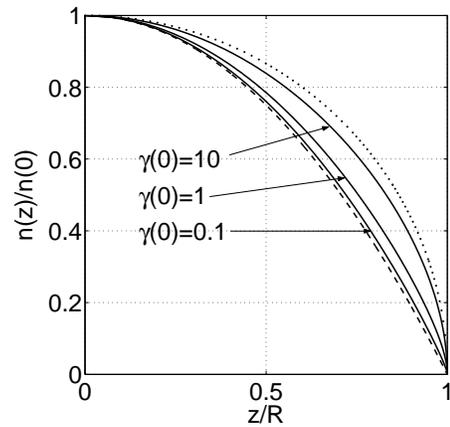}
\caption{Examples of the density profiles $n(z)/n(0)$ of a zero temperature $
1D$ Bose gas in a harmonic trap as a function of the dimensionless
coordinate $z/R$, for different values of the interaction parameter $\protect
\gamma(0)$. The solid lines are the results of the exact numerical solution
of the Lieb-Liniger equations within the LDA. The dashed and the dotted
lines are the analytic results given by the Thomas-Fermi parabola in the GP
regime and the square root of parabola in the TG regime, respectively.}
\label{density-zeroT}
\end{figure}

In Fig. \ref{density-zeroT} we present the density profiles $n(z)$ as a
function of the dimensionless coordinate $z/R$, for different values of $
\gamma(0)$. The full lines represent the results of the numerical
calculation within the LDA, which reproduce the results of Ref. \cite
{Olshanii2001}, while the dashed and the dotted lines represent,
respectively, the above analytic results in the GP and TG regimes.

\begin{figure}[ptb]
\includegraphics[width=6.5cm]{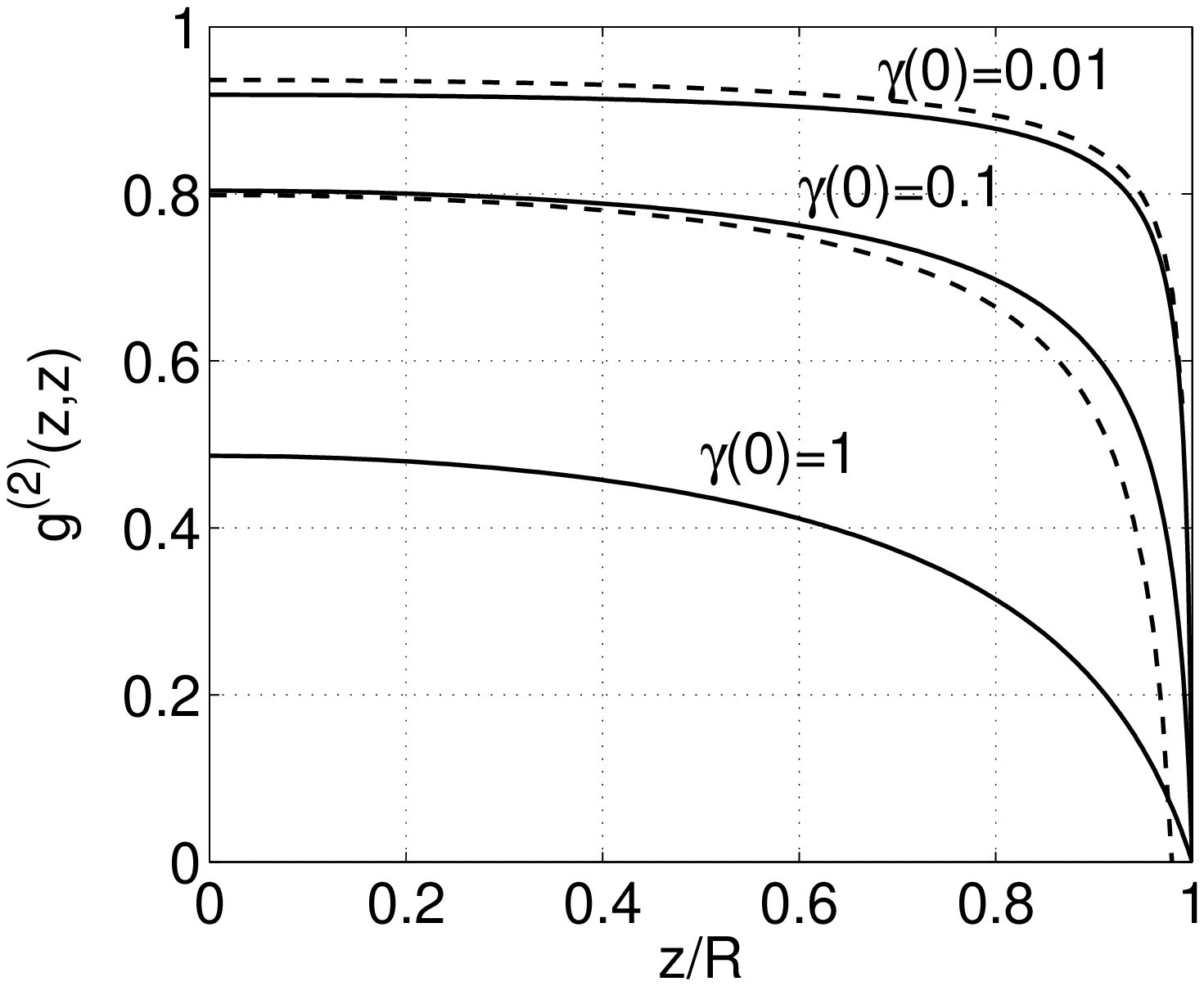}
\includegraphics[width=6.55cm]{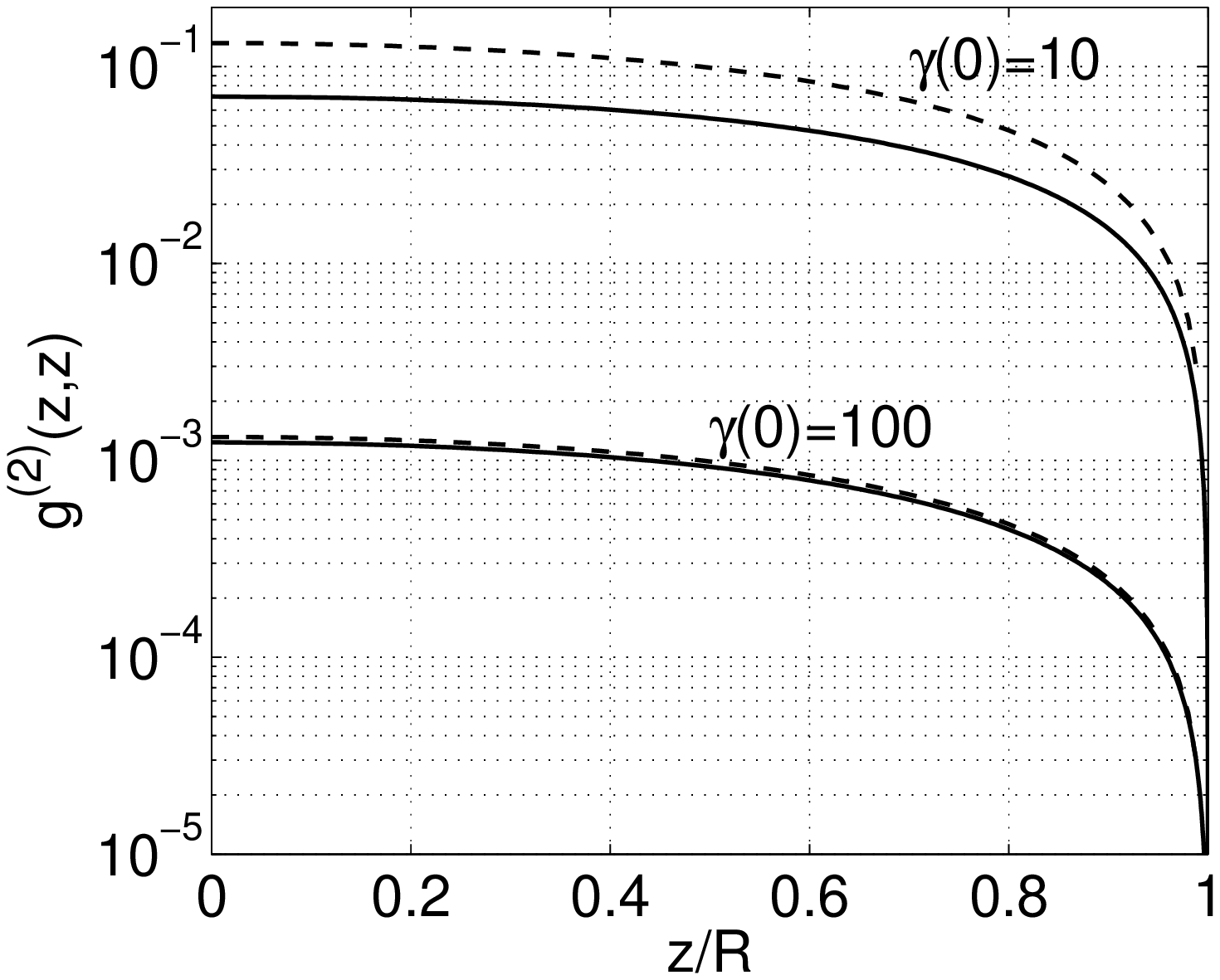}
\caption{The local pair correlation of a trapped $1D$ Bose gas at zero
temperature, $g^{(2)}(z,z)$, as a function of the displacement from the trap
center $z/R$, for different values of $\protect\gamma(0)$. The full lines
are the results of numerical calculation, while the dashed lines for $
\protect\gamma(0)=0.01,0.1$ and $\protect\gamma(0)=10,100$ are the
respective analytic results of Eqs. (\protect\ref{g2GP}) and (\protect\ref
{g2TG}) shown for comparison. }
\label{g2-zeroT}
\end{figure}

Figure \ref{g2-zeroT} shows the local pair correlation $g^{(2)}(z,z)$ as a
function of $z/R$, for different values of $\gamma(0)$. As we see, for the
case of weak interactions $\gamma(0)\ll1$, the pair correlation is close to
unity in the central bulk part of the distribution. This is an expected
result for the coherent or GP regime. As one approaches the tails of the
distribution, where the density is reduced and $\gamma(z)$ becomes larger
than one, the gas locally approaches the TG regime of `fermionization'.
Here, the pair correlation is suppressed below the coherent level, $
g^{(2)}(z,z)<1$.

For the cases where $\gamma(0)\gtrsim1$, including the TG regime of $\gamma
(0)\gg1$, the pair correlation is suppressed across the entire sample. In
the limit $\gamma(0)\rightarrow\infty$, the gas acquires pronounced
``fermionic'' properties so that the wave function strongly decreases as
particles approach each other, thus resulting in $g^{(2)}(z,z)\rightarrow0$.

\section{Finite temperature trapped gas}

\subsection{Key parameters}

An obvious choice of dimensionless interaction and temperature parameters
for describing a trapped $1D$ Bose gas within the LDA consists in using the
local value of the interaction parameter $\gamma(0)$ and the reduced
temperature $\tau(0)$ in the trap center.

These are the same parameters that are used in the uniform gas treatment
\cite{KK-DG-PD-GS-2003}, Eqs. (\ref{gamma}) and (\ref{tau1}), except that
now we define them via the local density $n(z)$. Thus, in general, we define
the local interaction parameter $\gamma (z)$ and the local reduced
temperature $\tau (z)$, according to:
\begin{align}
\gamma (z)& =\frac{mg}{\hbar ^{2}n(z)},  \label{tauz2} \\
\tau (z)& =\frac{T}{T_{d}(z)}=\frac{2mT}{\hbar ^{2}n^{2}(z)},
\end{align}
where $T_{d}(z)=\hbar ^{2}n^{2}(z)/(2m)$ is the local temperature of quantum
degeneracy that corresponds (locally) to the conditions where the mean
inter-particle separation becomes of the order of the thermal de Broglie
wavelength.

The values of these parameters at the trap center, $\gamma(0)$ and $\tau(0)$, 
completely characterize all relevant properties of the gas within the LDA,
including the associated density profiles $n(z)$, the resulting total number
of particles $N$, as well as the correlation functions and the thermodynamic
properties.

A completely equivalent pair of the interaction and temperature parameters,
which is, however, more suitable for practical purposes is the local value
of $\gamma(z)$ and a new temperature parameter $t$ defined via:
\begin{equation}
t=\frac{\tau(0)}{\gamma^{2}(0)}=\frac{\tau(z)}{\gamma^{2}(z)}=\frac{T} {
mg^{2}/(2\hbar^{2})}.  \label{t-def}
\end{equation}
According to this definition, the temperature is measured in units of the
characteristic energy $E_{b}=mg^{2}/(2\hbar^{2})$.

The advantage of using $t$ as the dimensionless temperature parameter is
that it is independent of the density and gives a direct measure of the
global temperature of the gas, which in equilibrium is the same for the
entire sample. This allows us to easily explore the
``interaction--temperature'' parameter space [$\gamma(0)-t$] in a systematic
way. For example, considering different values of $\gamma(0)$ while $t$ is
kept constant would correspond to physical conditions under which the peak
density of the gas $n(0)$ is varied while the absolute temperature $T$ is
kept unchanged. A novel experimental technique that implemented this
approach for achieving a Bose-Einstein condensation in a $3D$ gas has
recently been demonstrated in Ref. \cite{BEC-constT}.

Other alternative choices are possible for characterizing the interactions
and temperature of a trapped gas in dimensionless units. For example, to
characterize the system at different temperatures $T$ while the total number
of particles $N$ is kept constant, one can define an alternative pair of
global parameters which are more suitable for this case (see Sec. VI). Here,
the global temperature parameter can be defined as $\theta= T/T_{Q}$, where $
T_{Q}=N\hbar\omega_{z}$ is the global temperature of quantum degeneracy of a
trapped gas (in energy units, $k_{B}=1$, where $k_{B}$ is the Boltzmann
constant). Irrespective of the interaction strength, a harmonically trapped
Bose gas at $T\gg T_{Q}$ obeys the classical Boltzmann statistics, whereas
for $T\lesssim T_{Q}$ quantum statistical effects become important. This is
clearly seen in the limit of a trapped ideal gas ($\gamma(0)\rightarrow 0$)
and in the opposite limit of a strongly interacting gas ($
\gamma(0)\rightarrow \infty$). In the latter case, the problem maps onto the
trapped gas of non-interacting fermions \cite{Girardeau1960}. So, in both
limits $T_{Q}$ appears explicitly as the temperature of quantum degeneracy
for the trapped sample as a whole.

\subsection{LDA criterion at finite $T$}

Here we analyze the local density approximation for a finite temperature
gas, and obtain simple criteria for its validity in the limiting cases of
very high and very low temperatures, $T\gg T_{Q}$ and $T\ll T_{Q}$.

In the high temperature limit, $T\gg T_{Q}$ [in which case $\tau(0)\simeq
4\pi(T/T_{Q})^{2}$], the density distribution $n(z)$ can be approximated by
a Gaussian profile in all regimes, as the interaction between the particles
is negligible compared to their thermal kinetic energies. For $N$ particles
at temperature $T$ in a harmonic trap of frequency $\omega_{z}$, the density
profile is determined by the thermal distribution for a classical ideal gas
described by Boltzmann statistics:
\begin{equation}
n_{T}(z)=\frac{N}{\sqrt{\pi}R_{T}}\exp(-z^{2}/R_{T}^{2}),
\label{Thermal-Gaussian}
\end{equation}
where the radius $R_{T}$ characterizes the width of the Gaussian and is
given by
\begin{equation}
R_{T}=\sqrt{\frac{2T}{m\omega_{z}^{2}}.}  \label{R-T}
\end{equation}

In this high-temperature limit, the correlation length is given by the
thermal de Broglie wavelength $\Lambda_{T}$, so that the LDA criterion (\ref
{LDA-criterion}) gives:
\begin{equation}
z\ll\frac{T}{\hbar\omega_{z}}\left( \frac{2T}{m\omega_{z}^{2}}\right) ^{1/2}=
\frac{T}{\hbar\omega_{z}}R_{T},\;\;\;(T\gg T_{Q}).  \label{LDA-highT}
\end{equation}
Since $T\gg T_{Q}=N\hbar\omega_{z}$ implies that $T\gg\hbar\omega_{z}$, the
above LDA criterion can be satisfied for all locations $z$ from the trap
center up to distances equal to several characteristic widths $R_{T}$. For
sufficiently large total number of particles $N$, the ratio $T/\hbar
\omega_{z}$ will be even larger so that the LDA will be valid for even
larger distance from the trap center.

In the opposite limit of low temperatures $T\ll T_{Q}$ [in which case $
\tau(0)\simeq8\sqrt{2\gamma(0)}T/(3T_{Q})$ for $\gamma(0)\ll1$, and $
\tau(0)\simeq\pi^{2}T/T_{Q}$ for $\gamma(0)\gg1$], the density profile $n(z)$
can be approximated by two contributions. The first one is for the central
bulk part which will be close to the $T=0$ density profile up to a certain
distance $|z|\simeq R-\delta z$ ($\delta z\ll R$) from the trap center. Here
$R$ is the zero-temperature Thomas-Fermi radius, Eq. (\ref{R-TF-gen}). The
second contribution is for the tails of the distribution which can be
approximated by a thermal Gaussian.

As before, we will focus on the `weak' LDA condition (\ref{LDA-criterion})
involving the correlation length $l_{c}(z)$. For the central part of the
sample, up to distances $|z|\simeq R-\delta z$ ($\delta z\ll R$), we can use
the LDA criterion derived for the zero temperature gas, Eq. (\ref
{LDA-criterion-final}). In the weakly and strongly interacting limits this
can be rewritten -- as before -- in terms of the interaction parameter $
\gamma(0)$ and the total number of particles $N$ [see Eqs. (\ref{LDA-GP})
and (\ref{LDA-TG})].

For the Gaussian tails of the distribution, i.e. at distances $|z|>R$, where
the local correlation length is given by $\Lambda_{T}$, we use the above
high-temperature result, Eq. (\ref{LDA-highT}), and rewrite it in terms of
the Thomas-Fermi radius $R$ and the global chemical potential $\mu_{0}$. As
a result, the LDA criterion for the Gaussian tails of a low-temperature gas
can be written in the following form:
\begin{equation}
z\ll\left( \frac{T}{\hbar\omega_{z}}\right) ^{1/2}\left( \frac{\mu_{0} }{
\hbar\omega_{z}}\right) ^{1/2}R,\;\;\;(T\ll T_{Q}).  \label{LDA-tails}
\end{equation}

In the GP regime [$\gamma (0)\ll 1$], where $\mu _{0}\simeq gn(0)$, this
gives:
\begin{equation}
z/R\ll \tau (0)^{1/2}N.  \label{LDA-tails-GP}
\end{equation}
Thus, in order that the LDA works in the tails of the density distribution ($
|z|>R$) in the GP regime, one has to have $\tau (0)^{1/2}N\gg 1$. This
requirement can always be satisfied with a sufficiently large number of
particles $N$. For example, for $\tau (0)=3.8\times 10^{-3}$ [which can be
obtained, for example, with $\gamma (0)=0.01$ and $T/T_{Q}=0.01$, according
to the relationship $\tau (0)\simeq 8\sqrt{2\gamma (0)}T/(3T_{Q})$ valid in
this regime] one would need to have $N\geq 870$ in order to satisfy the LDA
criterion for distances $R\lesssim z\ll 20R$.

In the TG regime [$\gamma(0)\gg1$], where $\mu_{0}\simeq\pi^{2}\hbar
^{2}n^{2}(0)/(2m)$, Eq. (\ref{LDA-tails}) again reduces to the condition
given by Eq. (\ref{LDA-tails-GP}). Thus, in the TG regime the validity of
the LDA in the tails of the distribution again requires that $
\tau(0)^{1/2}N\gg1$. However, now we have $\tau(0)\simeq \pi^{2}T/T_{Q}$,
for $T\ll T_{Q}$.

To summarize, in the low temperature limit the LDA criterion can be easily
satisfied for the central bulk part of the density distribution and for the
Gaussian tails. This leaves the question of validity of the LDA in the
low-density region near $z=R$, where the density may vary more rapidly.

We note, however, that at small finite temperatures the variation of the
density profile around $z=R$ is more smooth than in the $T=0$ limit, so that
the LDA criterion may still be satisfied in this region, in contrast to the $
T=0$ case where the LDA necessarily breaks down as one approaches the edge
of the cloud at $z=R$. More importantly, the LDA becomes valid again for
distances past the small critical region around $z=R$, i.e. in the tails of
the density distribution. This means that the results of calculation of the
pair correlation function $g^{(2)}(z,z)$ at small finite temperatures should
be valid everywhere except in a small region around $z=R$. At high
temperatures the LDA criterion becomes less restrictive, and can be
satisfied for the entire sample.

\subsection{Calculating the local pair correlation and density profiles}

The local pair correlation $g^{(2)}(z,z)$, Eq. (\ref{g2z}), as a function of
the location $z$ from the trap center is calculated using Eq. (\ref{g2-grand}) 
in which $\mu$ is replaced by the local chemical potential $\mu(z)=\mu
_{0}-V(z)$ and where $n$ is the local density $n(z)$. The calculation is
based on iterative numerical solution of the Yang-Yang exact integral
equations \cite{Yang1969} for the excitation spectrum and for the
distribution function of ``quasi-momenta'', (\ref{YY1}) and (\ref{YY2}). For
a given set of values of $\mu(z)$, $T$ and $g$, this gives the resulting
density profile $n(z)$, Eq. (\ref{nzT}), and the pressure $P$, Eq. (\ref{P}). Differentiating $P$ with respect to $g$ gives the local pair correlation 
$g^{(2)}(z,z)$.

A convenient way to implement the numerical algorithm for solving the
Yang-Yang equations is via a dimensionless coordinate
\begin{equation}
\xi=\frac{z}{R_{T}},  \label{xi}
\end{equation}
where the length scale $R_{T}$ is the thermal width of the classical
Gaussian distribution $n_{T}(z)$, given by Eq. (\ref{R-T}).

Using the dimensionless coordinate $\xi$, the local chemical potential can
be rewritten as
\begin{equation}
\mu(R_{T}\xi)=\mu_{0}-T\xi^{2}.  \label{mu-local}
\end{equation}
After setting up a lattice of $\xi$ values, \{$\xi_{i}$\}, the solution to
the Yang-Yang equations proceeds as in the case of a uniform gas, with the
input parameters being an array of the values of the local chemical
potential $\mu_{i}=\mu(R_{T}\xi_{i})$, the temperature $T$, and the coupling
parameter $g$, as described above.

The final numerical results are then presented in terms of the dimensionless
parameters $\gamma(0)$ and the temperature parameter $\tau(0)$ (or $t$),
where we note that
\begin{equation}
\frac{n(z)}{n(0)}=\frac{\gamma(0)}{\gamma(z)}.  \label{nzovern0}
\end{equation}
This makes the output results scalable with respect to the physical
parameters, rather than dependent of their absolute values.

The total number of particles in the system is calculated from the resulting
density profile $n(z)$ via
\begin{equation}
N=\int n(z)dz.  \label{N}
\end{equation}

Using the dimensionless coordinate $\xi$ and Eq. (\ref{nzovern0}) this can
be rewritten as
\begin{equation}
N=\frac{R_{T}mg}{\hbar^{2}\gamma(0)}\int_{-\infty}^{+\infty}\frac{\gamma (0)
}{\gamma(R_{T}\xi)}d\xi,
\end{equation}
so that the dimensionless ratio $T_Q/T$ is given by
\begin{equation}
\frac{T_Q}{T}=\frac{2}{\sqrt{t}}\int_{-\infty }^{+\infty}\frac{d\xi}{
\gamma(R_{T}\xi)}.
\end{equation}
This gives a relationship between the global and local dimensionless
parameters and allows us to present the final results in a scalable fashion,
rather than in terms of the absolute values of $N$, $T$, $\omega_{z}$, and $
g $. Here, the desired values of $T_Q/T$ can be achieved by varying the
ratio $\mu_{0}/T$ of the input parameters $\mu_{0}$ and $T$.

\section{Density profiles and pair correlations}

\subsection{Regimes in a uniform gas}

\label{sec:analpred}In order to understand the results for the pair
correlations $g^{(2)}(z,z)$ of a trapped $1D$ Bose gas, we first recall the
classification of the regimes of a \emph{uniform} gas. In Ref. \cite
{KK-DG-PD-GS-2003}, these were identified using the results for the local
pair correlation $g^{(2)}$ in terms of the interaction parameter $\gamma$
and the reduced temperature $\tau$. Here, we give a brief summary of these
results, except that we rewrite them in terms of the parameters $\gamma$ and
$t=\tau/\gamma^{2}$, instead of $\gamma$ and $\tau$. This is completely
equivalent to the original pair. The new parameters $\gamma$ and $t$ are
more suitable for exploring the properties of the \textit{trapped} gases, as
discussed in Sec. IV A.

\begin{figure}[ptb]
\centering
\includegraphics[width=6.5cm]{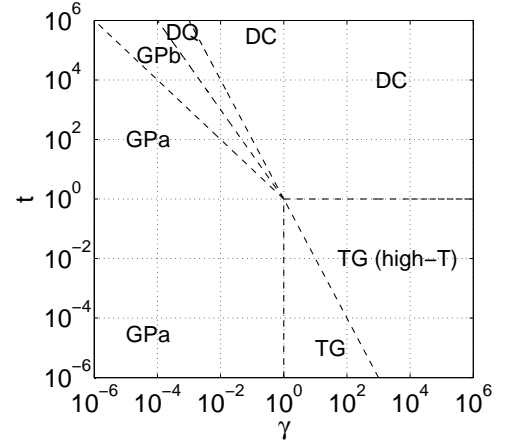}
\caption{Diagram of different regimes of a uniform $1D$ Bose gas in the ($\protect
\gamma$-$t$) plane. The labels TG, GP, DQ and DC refer to the
Tonks-Girardeau, Gross-Pitaevskii, decoherent quantum and decoherent
classical regimes, respectively. Although all transitions are continuous, for purposes
of discussion we
classify the distinct regimes  as follows:\\
TG: $\gamma>1$, $t<\gamma^{-2}$;
TG(high-T): $\gamma^{-2}<t<1$; \\
GPa: $\gamma<1$, $t<\gamma^{-1}$;  
GPb: $\gamma^{-1}<t<\gamma^{-3/2}$; \\
DQ:  $\gamma^{-3/2}<t<\gamma^{-2}$; 
DC: $t>max\{1, \gamma^{-2}\}$. 
}
\label{regionstgamma}
\end{figure}

The diagram representing these different regimes for a uniform 1D Bose gas
in the parameter space ($\gamma$-$t$) is shown in Fig. \ref{regionstgamma}.
The regimes are classified as follows:

\emph{Strong coupling regime}: In the strong coupling TG regime of `fermionization',
where $\gamma \gg 1$ and the temperature $T\ll T_{d}$ ($\tau \ll 1$ or $t\ll \gamma ^{-2}$),
we only have a small correction compared to the zero temperature result \cite{KK-DG-PD-GS-2003,Cazalilla-PRA}:
\begin{equation}
g^{(2)}\simeq \frac{4}{3}\left( \frac{\pi }{\gamma }\right) ^{2}\left[ 1+
\frac{\gamma ^{4}t^{2}}{4\pi ^{2}}\right];\,\,\,t\ll \gamma ^{-2},\,\gamma
\gg 1;\,\,\,\text{TG}.  \label{TG-low-T}
\end{equation}

In the case of strongly interacting nondegenerate bosons, where $\gamma \gg
1 $ and the temperature $T\gg T_{d}$ ($1\ll \tau \ll \gamma ^{2}$ or $\gamma
^{-2}\ll t\ll 1$), we have the regime of high-temperature `fermionization'.
Despite the temperature $T\gg T_{d}$, the local pair correlation is strongly suppressed
($g^{(2)}\ll 1$) \cite{KK-DG-PD-GS-2003,Cazalilla-PRA}:
\begin{equation}
g^{(2)}\simeq 2t;\;\;\;\;\gamma ^{-2}\ll t\ll 1;\,\,\text{TG (high-T)}.
  \label{TG-high-T}
\end{equation}

\emph{Gross-Pitaevskii regime:} In the GP regime, where $\gamma \ll 1$,
the chemical potential is $\mu =ng$ and
at temperatures $T\ll \mu =2T_{d}\gamma $ ($\tau \ll \gamma $ or $t\ll
\gamma ^{-1}$) the finite temperature correction to the zero temperature
result is again very small \cite{KK-DG-PD-GS-2003}:
\begin{equation}
g^{(2)}\simeq 1-\frac{2}{\pi }\sqrt{\gamma }+\frac{\pi }{24}t^{2}\gamma
^{5/2};\;\,\gamma \ll 1,\ t\ll \gamma ^{-1};\;\,\text{GPa}.
\label{GPa}
\end{equation}

For $T\gg \mu =2T_{d}\gamma $ ($\tau \gg \gamma $ or $t\gg \gamma ^{-1}$),
the finite temperature correction is the leading one. It is important to
recognize that the upper bound for the GP regime extends only up to
temperatures of the order of $T\sim \sqrt{\gamma }T_{d}$ ($\tau \sim \sqrt{
\gamma }$ or $t\sim \gamma ^{-3/2}$), and not to $T\sim T_{d}$. Here, the
temperature $T_{d}$ is responsible for the presence of the quantum
degeneracy, while $\sqrt{\gamma }T_{d}$ for the presence of phase
coherence. Thus, for $T\gg \mu $, the finite-temperature GP regime lies
within the temperature interval $2T_{d}\gamma \ll T\ll \sqrt{\gamma }T_{d}$
($\gamma \ll \tau \ll \sqrt{\gamma }$ or $\gamma ^{-1}\ll t\ll \gamma ^{-3/2}$),
and the pair correlation here is given by \cite{KK-DG-PD-GS-2003}:
\begin{equation}
g^{(2)}\simeq 1+\frac{1}{2}t\gamma ^{3/2};\;\;\ \;\gamma ^{-1}\ll
t\ll \gamma ^{-3/2};\,\;\,\text{GPb}.  \label{GPb}
\end{equation}

\emph{Decoherent regime:} $t\gg \max (1,\gamma ^{-3/2})$. Due to the
existence of two characteristic temperatures in the $1D$ uniform gas, $\sqrt{
\gamma }T_{d}$ and $T_{d}$, at temperatures $T$ higher than $\sqrt{\gamma }
T_{d}$ one has two sub-regions. For temperatures in the
interval $\sqrt{\gamma }T_{d}\ll T\ll T_{d}$ ($\sqrt{\gamma }\ll \tau \ll 1$
or $\gamma ^{-3/2}\ll t\ll \gamma ^{-2}$) the gas is in the decoherent
quantum (DQ) regime, while for $T\gg T_{d}$ ($\tau \gg 1$ or $t\gg \gamma
^{-2}$) the gas is in the decoherent classical (DC) regime. In both cases
the local pair correlation is close to $g^{(2)}\simeq 2$ \cite
{KK-DG-PD-GS-2003}:
\begin{equation}
g^{(2)}\simeq 2-4/(t^{2}\gamma^{3});\;\;\;\gamma ^{-3/2}\ll t\ll \gamma ^{-2};\,\,\,\text{DQ},  \label{DQ-eq}
\end{equation}
\begin{equation}
g^{(2)}\simeq 2-\left( \frac{2\pi }{t}\right) ^{1/2};\;\;\;t\gg \max
(\gamma ^{-2},1);\,\,\,\text{DC}.  \label{DC-eq}
\end{equation}
The result in the DC regime remains valid for large $\gamma $, provided $
\tau \gg \gamma ^{2}$ (or $t\gg \gamma ^{-2}$) \cite{KK-DG-PD-GS-2003}, and
we can combine the required conditions on temperature via $t\gg \max (\gamma
^{-2},1)$.

\subsection{Regimes in a trapped gas}

In a harmonically trapped finite-temperature $1D$ Bose gas we again have a
strong coupling regime, weak coupling GP regime, and a decoherent regime.
The results for the local pair correlation $g^{(2)}(z,z)$ in the first two
regimes are easily obtained from Eqs. (\ref{TG-low-T})-(\ref{GPb}) by
replacing the interaction parameter $\gamma $ by the local $z$-dependent
value $\gamma (z)$ of the trapped sample. However, it is convenient to rewrite the
results for the local correlation $g^{(2)}(0,0)$ in the trap center in terms
of $\gamma (0)$ and the temperature parameter $\theta \equiv T/T_{Q}$, where
$T_{Q}=N\hbar \omega _{z}$ is the global temperature of quantum degeneracy
of the sample as a whole.

\emph{Strong coupling regime.} In the strong coupling
TG regime, where $\gamma (0)\gg 1$ and $T\ll T_{Q}$,
the density profile is given by the Thomas-Fermi result, Eq. (\ref
{TF-parabola-TG}), and this allows one to establish the relationship between
the temperature parameters $t$ and $\theta $, using $\tau (0)\simeq \pi
^{2}T/T_{Q}$ valid in this regime. Thus, $t\simeq \pi ^{2}\theta /\gamma
^{2}(0)$ and Eq. (\ref{TG-low-T}) transforms into:
\begin{equation}
g^{(2)}(0,0)\simeq \frac{4\pi ^{2}}{3\gamma ^{2}(0)}\left[ 1+\frac{\pi ^{2}}{
4}\theta ^{2}\right] ,  \label{TG-low-T-global}
\end{equation}
where $\theta \ll 1/\pi ^{2}$, and $\gamma (0)\gg 1$.

For the regime of high-temperature fermionization at $T\gg T_{Q}$, the
density profile is given by the thermal Gaussian, Eq. (\ref{Thermal-Gaussian}
), so that $\tau (0)\simeq 4\pi (T/T_{Q})^{2}$ and hence $t\simeq 4\pi
\theta ^{2}/\gamma ^{2}(0)$. Therefore, Eq. (\ref{TG-high-T}) transforms into:
\begin{equation}
g^{(2)}(0,0)=\frac{8\pi }{\gamma ^{2}(0)}\theta ^{2},
\label{TG-high-T-global}
\end{equation}
where $1\ll \theta \ll \gamma (0)/\sqrt{4\pi}$.

\emph{Gross-Pitaevskii regime.} In the GP regime
($\gamma (0)\ll 1$), for temperatures $T\ll T_{Q}$, the density
profile is given by the Thomas-Fermi inverted parabola, Eq. (\ref
{TF-parabola}). We then have $\tau (0)\simeq 8\sqrt{2\gamma (0)}T/(3T_{Q})$
or $t\simeq 8\sqrt{2}\theta \lbrack \gamma (0)]^{-3/2}/3$, so that Eq. (\ref
{GPa}) transforms into
\begin{equation}
g^{(2)}(0,0)=1-\frac{2}{\pi }\sqrt{\gamma (0)}+\frac{16\pi }{27\sqrt{\gamma
(0)}}\theta ^{2},  \label{GPa-global}
\end{equation}
where $\theta \ll 3\sqrt{\gamma (0)}/(8\sqrt{2})$ ($\theta \ll
0.27\sqrt{\gamma (0)}$) and $\gamma (0)\ll 1$. Similarly, Eq.
(\ref{GPb}) transforms into
\begin{equation}
g^{(2)}(0,0)=1+\frac{4\sqrt{2}}{3}\theta ,  \label{GPb-global}
\end{equation}
where $3\sqrt{\gamma (0)}/(8\sqrt{2})\ll \theta \ll 3\sqrt{2}/8$
($0.27\sqrt{\gamma (0)}\ll \theta \ll 0.27$).

It is important to emphasize that for small $\gamma (0)$ and making the temperature
sufficiently low ($T/T_Q = \theta \ll 1$)
the gas is always in the Gross-Pitaevskii regime. 

\emph{Decoherent regime.} For $T\gg T_{Q}$, a harmonically
trapped $1D$ Bose gas is in the decoherent
classical regime. Here, the density profile is given by the thermal
Gaussian, Eq. (\ref{Thermal-Gaussian}), so that $\tau (0)\simeq 4\pi
(T/T_{Q})^{2}$ and hence $t\simeq 4\pi \theta ^{2}/\gamma ^{2}(0)$.
Accordingly, Eq. (\ref{DC-eq}) transforms into:
\begin{equation}
g^{(2)}=2-\frac{\gamma (0)}{\sqrt{2}}\theta ,  \label{DC-eq-global}
\end{equation}
where $\theta \gg \max \{1,\gamma (0)\}$, neglecting a numerical factor of
the order of one. Thus, the validity of this result requires $\theta \gg 1$
for $\gamma (0)<1$, and $\theta \gg \gamma (0)$ for $\gamma (0)>1$.

\begin{figure}[tbp]
\includegraphics[width=5.7cm]{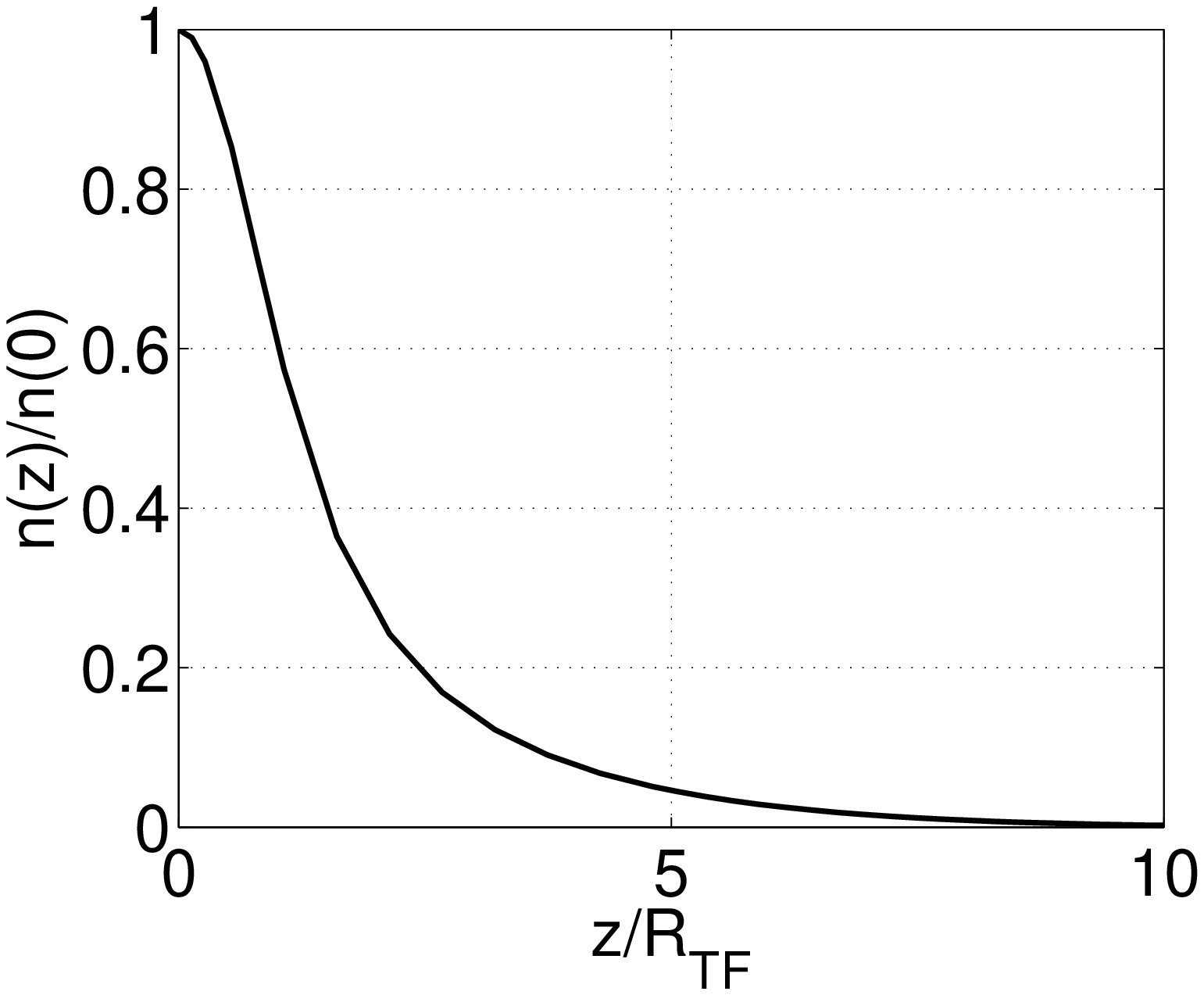}
\includegraphics[width=5.7cm]{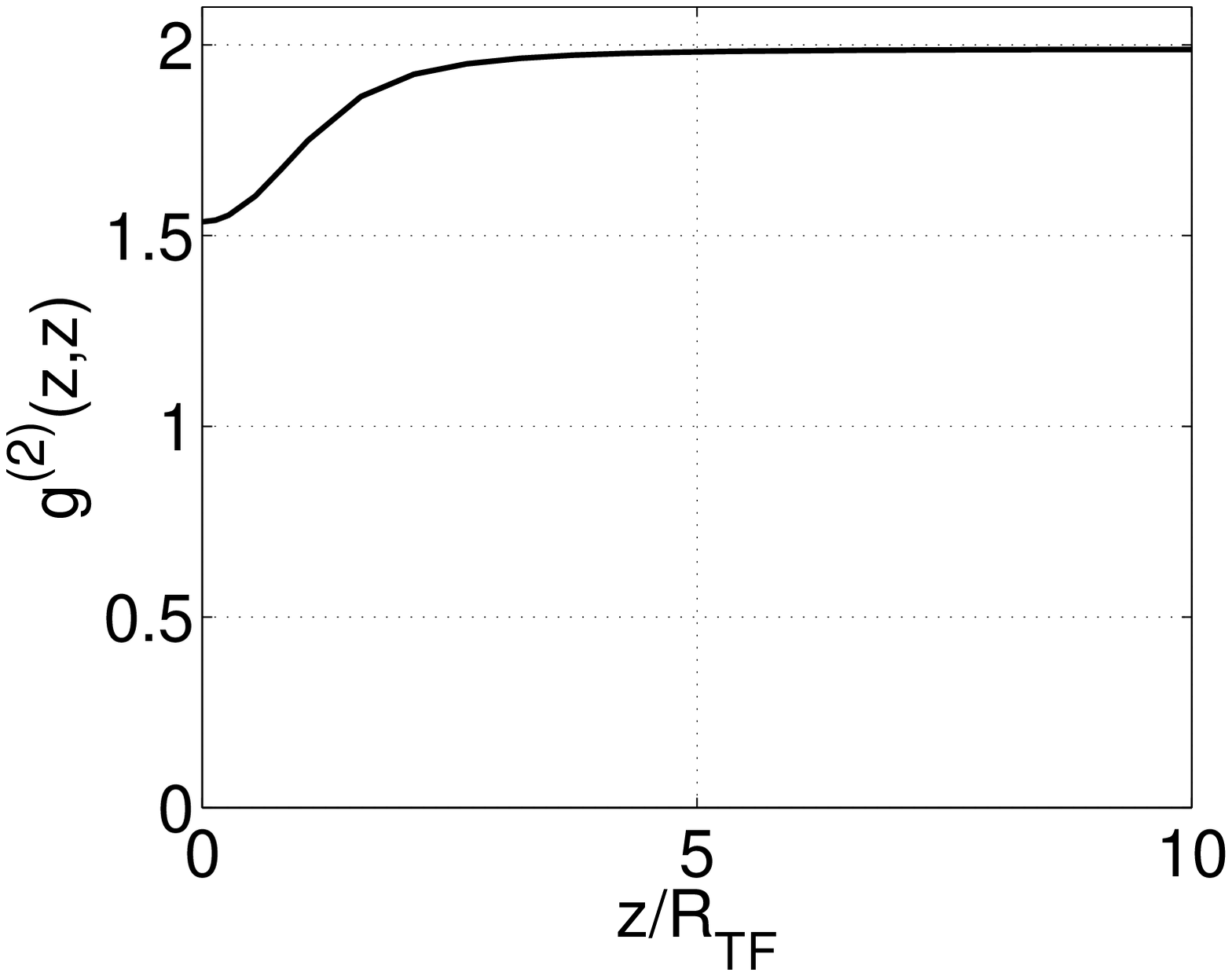}
\caption{Density profile $n(z)/n(0)$ and the local pair correlation $
g^{(2)}(z,z)$ as a function of the distance from the trap center $z/R_{TF}$,
for $T=0.2T_{Q}$ ($t=5\times 10^{4}$) and $\protect\gamma (0)=1.14\times
10^{-3}$. This corresponds to case 1b in the diagram of Fig. \protect\ref
{linesofconstt} below.}
\label{DQ}
\end{figure}

In the case of small $\gamma (0)$ and at temperatures
$ T\lesssim T_{Q}$, one
has a crossover from the classical decoherent ($T\gg T_{Q}$) regime to the
GP ($T\ll T_{Q}$) regime. The properties of the gas in this region can be
treated as containing (locally) features of the decoherent quantum regime of
the uniform gas. An example illustrating this behaviour is given in Fig. \ref
{DQ} where we plot the density profile $n(z)$ and the local pair correlation
$g^{(2)}(z,z)$ as a function of $z/R_{TF}$ where $R_{TF}$ is the
Thomas-Fermi radius in the GP regime. These are calculated numerically using
the solution to the Yang-Yang integral equations, with a value of $\gamma(0)=1.14 \times 10^{-3}$ 
at the trap center. 

The temperature in this
example is  $T=0.2T_{Q}$, which is intermediate between the
decoherent classical and the GP regimes. Locally, the tails of the density profile are in
the decoherent classical regime. On the other hand, the central part has features of the
decoherent quantum regime. The figure for $g^{(2)}(z,z)$ shows that
fluctuations well above the coherent level, with $g^{(2)}(0,0)\simeq 1.5$,
can occur even at temperatures below the transition to a quantum gas.
However, with further temperature reduction well below $T_{Q}$, the density
profile shrinks and one always has a coherent GP regime. For smaller
values of $\gamma$, the temperature at which the coherent regime emerges
must in general be calculated numerically.

\subsection{Variation with temperature}

\begin{figure}[ptb]
\includegraphics[width=6cm]{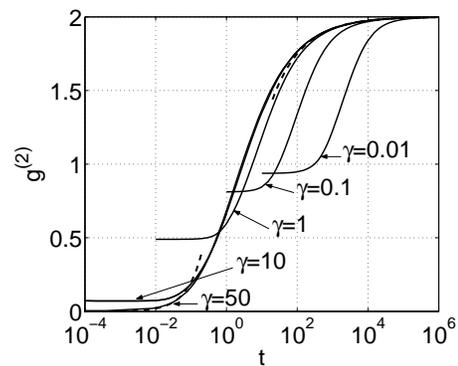}
\caption{Local pair correlation $g^{(2)}$ as a function of the reduced
temperature $t$, for different values of the interaction parameter $\protect
\gamma$. The solid lines are the exact numerical results, while the dashed
lines correspond to the approximate analytic result of Eq. (\protect\ref
{eq:g2edge}). }
\label{g2edgepic}
\end{figure}

In Fig. \ref{g2edgepic}, we illustrate different regimes by plotting the
local pair correlation $g^{(2)}$ as a function of the temperature parameter $
t$, for different values of $\gamma $.

For sufficiently large $\gamma$ the pair correlation approaches a universal function of
the parameter $t$:
\begin{equation}
\lim_{\gamma\rightarrow\infty}g^{(2)}=\left\{
\begin{array}{ll}
2t\;, & t\ll1, \\
2-\sqrt{2\pi/t}\;, & t\gg1.
\end{array}
\right.  \label{eq:g2edge}
\end{equation}
This is because by increasing $\gamma$ one can always reach
locally the condition $t\gg \gamma^{-2}$. Then, for $t\ll 1$ one
has (locally) the regime of high-temperature fermionization and
can use Eq.(\ref{TG-high-T}), whereas for $t\gg 1$ the sample will
be in the decoherent classical regime described by
Eq.(\ref{DC-eq}).

This has an interesting consequence at sufficiently low temperatures
$t\ll 1$. For $\gamma \gg 1$, which is always the case for far
tails of the density distribution, the pair correlation remains suppressed
below the coherent level ($g^{(2)}<1$) rather than approaches the value of $
g^{(2)}\simeq2$. This occurs despite the gas is locally not
quantum degenerate at low density. One thus sees that fermionization in which the Bose gas
develops antibunching with $g^{(2)}\rightarrow 0$ is an explicitly
low-temperature phenomenon, when the temperature is scaled relative to the
interaction strength.

However, for $\gamma\gg 1$ the suppression of pair correlations is not
temperature-independent. Instead, the numerical results for increasing $
\gamma$ converge to a single universal function of $t$.

\subsection{Spatial variation}

In the case of a trapped gas, the same diagram of Fig. \ref{regionstgamma}
also describes the spatial variation of the gas within the LDA. The
parameter $\gamma$ now becomes position dependent, $\gamma(z)$, due to the
dependence on the density $n(z)$. In this diagram, any point on the ($\gamma
$-$t$) plane can be thought of as representing the interaction parameter $
\gamma(0)$ evaluated at the trap center, and the dimensionless temperature $
t $. This is sufficient to completely characterize the properties of the
trapped gas. The subsequent local values of $\gamma(z)$ of such a gas -- as
one moves from the trap center towards the tails of the density distribution
-- can be represented by a horizontal line drawn in the direction of
increasing $\gamma(z)$ at constant $t$.

\begin{figure}[ptb]
\includegraphics[width=6cm]{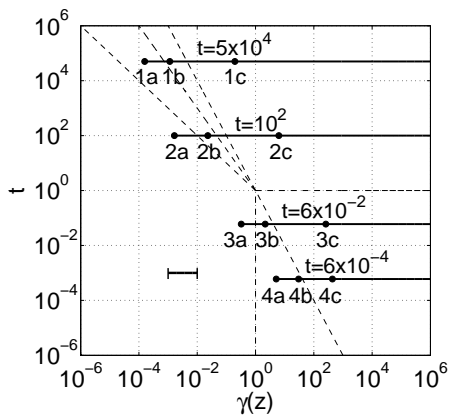}
\caption{Same as in Fig. \protect\ref{regionstgamma}, except with four
horizontal lines at different temperatures $t$. This explores different
density profiles in the parameter space [$\protect\gamma(0)$-$t$], where the
points 1a-1c, 2a-2c, 3a-3c, and 4a-4c (marked by circles) are representative
examples corresponding to different values of $\protect\gamma(0)$ at the
trap center, at different temperatures $t$. }
\label{linesofconstt}
\end{figure}

This is shown in Fig. \ref{linesofconstt}, where the four horizontal lines
correspond to four different temperatures $t$, while various points along
each line represent different `initial' values of the interaction parameter $
\gamma(0)$. The interval in the left lower corner of the diagram shows the
displacement (notice the logarithmic scale) for which the local value of $
\gamma(z)$ is increased by a factor of $10$. This corresponds to a $10$ fold
decrease in the density $n(z)$. For any given distribution with the value of
$\gamma(0)$ in the center, this interval helps to immediately determine what
fraction of the density profile relative to the peak density $n(0)$ is
contained within a certain regime.

In the very far tails ($z\rightarrow\infty$) of any density distribution,
where $n(z)$ vanishes and $\gamma(z)\rightarrow\infty$, we always enter
either the DC or the high-temperature TG regime, depending on the
temperature $t$. In addition, by considering a sample at any fixed
temperature $t$, while the peak density is increased [$\gamma(0)$ is
decreased], one can always reach the situation where the bulk of the density
distribution is in the GP regime where $g^{(2)}(z,z)\simeq1$. Physically,
this can be achieved by adding more particles to the system while
maintaining the same global temperature $T$, under constant coupling $g$.
From Fig. \ref{g2edgepic} it is clear that the density required may be
relatively high, with $\gamma(0)=0.01$ being necessary to have a limiting
value of $g^{(2)}\ge 0.9 $ at $t=10^2$, as an example. For $\gamma(0)>1$,
there is no coherent GP regime over the entire range of temperatures.

To illustrate different examples, we now calculate the density profiles $
n(z)/n(0)$ and the local pair correlations $g^{(2)}(z,z)$ as a function of
the distance from the trap center $z$. The distance $z$ is conveniently
plotted in units of the Thomas-Fermi radius in the GP regime, $R_{TF}$,
given by by Eq. (\ref{TF-GP-radius}). The relationship between $z/R_{TF}$
and the dimensionless coordinate $\xi=z/R_{T}$ is: $z/R_{TF}=(\sqrt {2}
/4)^{1/3}t^{1/2}\gamma(0)^{1/2}\xi$. For a gas with a given coupling
constant $g$, the temperature parameter $t$ gives the measure of the
absolute temperature $T$. In this sense, the examples with $t>1$ and $t<1$
in Fig. \ref{linesofconstt} represent high- and low-temperature limits,
which we analyze separately.

\subsubsection{High-temperature case}

\begin{figure}[ptb]
\centering
\includegraphics[height=7.5cm]{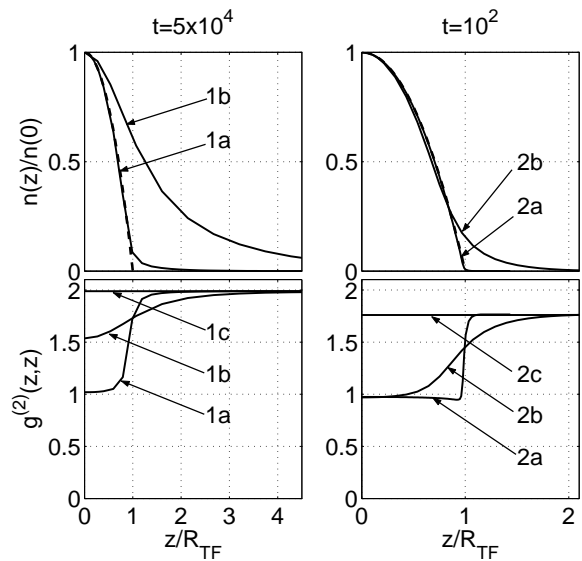}
\caption{Density profiles $n(z)/n(0)$ and the local pair
correlation $g^{(2)}(z,z)$ as a function of $z/R_{TF}$ for a
harmonically trapped $1D$ Bose at different temperatures $t$:
$t=5\times10^{4}$ (first column) and $t=10^{2}$ (second column).
The values of the interaction parameter $\gamma(0)$ in the trap
center for each of the curves are as follows: 1a --
$\gamma(0)=1.57\times10^{-4}$, 1b --
$\gamma(0)=1.14\times10^{-3}$, 1c -- $\gamma(0)=0.196$, 2a --
$\gamma(0)=1.65\times10^{-3}$, 2b --
$\gamma(0)=2.31\times10^{-2}$, and 2c -- $\gamma(0)=6.30$. The dashed
lines represent
the Thomas-Fermi inverted parabola, Eq. (\ref{TF-parabola}). The
density profiles corresponding to the lower density cases 1c, and
2c are well approximated by the thermal (Gaussian) distribution
for a classical ideal gas, Eq. (\ref{Thermal-Gaussian}), and are
omitted from the graphs for clarity. The respective pair
correlations for these low density cases are almost constant along
the entire sample and are given by the value of $g^{(2)}(z,z)$ in
the tails of the distribution $z\rightarrow\infty$. Depending on
the temperature $t$, these values can be determined using the
results of Fig. \ref{g2edgepic} at $\gamma(z)\gg1$ [see the curve
for $\gamma(z)=50$]. } \label{nz-g2-12}
\end{figure}

In Fig. \ref{nz-g2-12} we present examples of calculated density profiles $
n(z)$ and local pair correlations $g^{(2)}(z,z)$, for the high-temperature
cases of $t=5\times10^{4}$ and $t=10^{2}$. The examples shown correspond to
the points marked by circles 1a-1c and 2a-2c in the diagram of Fig. \ref
{linesofconstt}. For each temperature $t$, the sequence of points 1a, 1b and
1c (and similarly for 2a, 2b and 2c) corresponds to a decreasing peak
density of the gas [increasing values of $\gamma(0)$], while the absolute
temperature $T$ is kept constant. This can be achieved by decreasing the
total number of particles $N$ in the sample, at constant $T$.

The examples 1c and 2c represent a low-density (non-degenerate) gas in the
decoherent classical (DC) regime. The corresponding density profiles are
well approximated by a thermal Gaussian, Eq. (\ref{Thermal-Gaussian}), and
are omitted from the graphs for clarity. The respective second-order
correlation functions $g^{(2)}(z,z)$, display large thermal (Gaussian)
density fluctuations with $g^{(2)}(z,z)\simeq2$.

Moving along the horizontal lines in the direction of decreasing $\gamma(0)$
(starting from the points 1c or 2c for each temperature $t$) corresponds to
increasing peak densities of the gas. As a result one crosses the respective
boundaries and enters different regimes of quantum degeneracy shown in Fig.
\ref{regionstgamma}. Here, the limiting regime as $\gamma(0)\rightarrow0$ at
constant $t$ is the GP regime where the density profiles are well
approximated by the Thomas-Fermi parabola (\ref{TF-parabola}) (see the
graphs corresponding to 1a and 2a), while the pair correlation in the bulk
of the density profile is close to that of the coherent level $
g^{(2)}(z,z)\simeq1$.

The intermediate values of $\gamma(0)$ are represented by the examples 1b
and 2b which have density profiles that are intermediate between a Gaussian
and the inverted parabola. The respective pair correlations $g^{(2)}(z,z)$
also take intermediate values $1\lesssim g^{(2)}(z,z)<2$. In the example 2b,
however, the central part of the density profile is in the GP regime, so
that the departures form the coherent level of fluctuations are only seen in
the tails of the density profiles.

\subsubsection{Low-temperature case}

\begin{figure}[ptb]
\centering
\includegraphics[height=7.5cm]{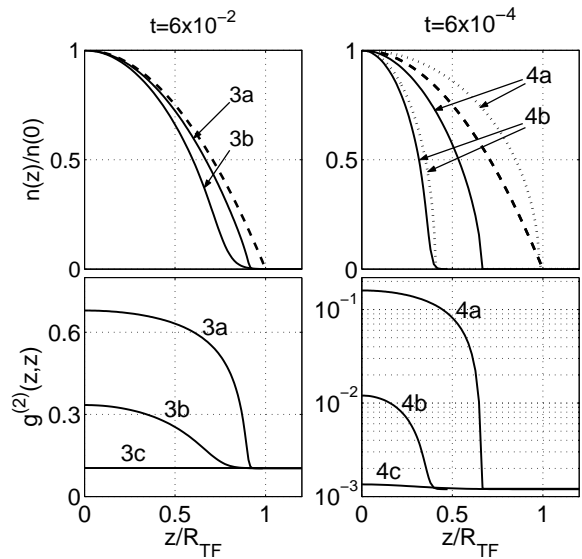}
\caption{Same as in Fig. \ref{nz-g2-12}, except for:
$t=6\times10^{-2}$ (first column) and $t=6\times10^{-4}$ (second
column). The values of the interaction parameter $\gamma(0)$ in
the trap center for each of the curves are as follows: 3a --
$\gamma(0)=0.323$, 3b -- $\gamma(0)=2.16$, 3c --
$\gamma(0)=2.58\times10^{2}$, 4a -- $\gamma(0)=5.13$, 4b --
$\gamma(0)=30.0$, and 4c -- $\gamma(0)=4.35\times10^{2}$.
The dashed lines represent
the Thomas-Fermi inverted parabola in the GP regime, Eq. (\ref{TF-parabola}),
while the dotted lines correspond to the Thomas-Fermi square
root of parabola in the TG regime, Eq. (\ref{TF-parabola-TG}).
}
\label{nz-g2-34}
\end{figure}

Next, we consider the low-temperature behaviour, in which evidence for the
Tonks-Girardeau `fermionization' can occur. Fig. \ref{nz-g2-34} represents
examples of the density profiles and pair correlations for a gas with lower
values of the temperature parameter $t$: $t=6\times10^{-2}$ and $
t=6\times10^{-4}$. The examples shown correspond to the points 3a-3c and
4a-4c in the diagram of Fig. \ref{linesofconstt}. As wee see, for $
\gamma(0)>1$ the gas is in the Tonks-Girardeau regime. Comparing this with
the earlier high-temperature examples, we see that for a given density with $
\gamma(0)>1$, achieving the Tonks-Girardeau regime requires lower
temperatures, $t<1$.

Here again, the low-density [large $\gamma(0)$] examples of 3c and 4c have
density profiles that are well approximated by the thermal Gaussian, Eq. (
\ref{Thermal-Gaussian}), and are omitted from the graphs for clarity.
However, the pair correlations do not display large thermal fluctuations,
but rather are suppressed below the coherent level, $g^{(2)}(z,z)<1$. This
reflects the fact that the gas is in the regime of high-temperature
``fermionization''. The example 4c corresponds to the lowest temperature
parameter $t$, which at constant peak density [or constant $\gamma(0)$]
corresponds to the largest interaction strength $g$, hence the smallest
value of $g^{(2)}(z,z)$.

The example 4b is deep enough in the TG regime, and we see that the density
profile is close to the Thomas-Fermi square root of parabola, Eq. (\ref
{TF-parabola-TG}), while the pair correlation is well below the coherent
level $g^{(2)}(z,z)\ll1$. The example 4a is for a smaller value of $
\gamma(0) $ [higher peak density], which is closer to the boundary with the
GP regime. As a result, the shape of the density profile departs from the
respective TG result and is intermediate between the TG and GP parabolas,
while the pair correlation in the central part of the density distribution
increases. Finally, the examples 3b and 3a are for intermediate values of $t$
and $\gamma(0)$ which are not well described by analytical approaches.

In all these examples the limiting behaviour of the pair correlation $
g^{(2)}(z,z)$ in the far tails of the density distribution is described by a
universal function of $t$, as discussed earlier (see Fig. \ref{g2edgepic}).
The overall conclusion that can be drawn from this analysis is that the
local pair correlation $g^{(2)}(z,z)$ can vary between a broad range of
values between zero and two and has a rich built-in structure. It
provides far more sensitive information about the regimes of trapped $1D$
Bose gases than the respective density profiles.
\ \newline

\section{Trapped gas at constant $N$}

Here we investigate the properties of a trapped gas at different
temperatures $T$ and constant total number of particles $N$. Since the
overall picture in terms of the density profiles and the behaviour of the
local pair correlation has already been understood in terms of the diagram
of Figs. \ref{regionstgamma} and \ref{linesofconstt}, it is now sufficient
to only monitor the changes in the temperature parameter $t$ and the value
of the interaction parameter $\gamma(0)$ under conditions when $N$ is kept
constant, and then map these changes into the [$t$-$\gamma(0)$] plane.

Thus, for a given system with the coupling $g$, trap frequency $\omega_{z}$,
and the total number of particles $N$, our task consists of calculating the
density profiles $n(z)$ at different temperatures $T$, with the constraint
that the total number of particles remains unchanged. Once this is done, we
identify the respective values of the dimensionless temperature parameter $t$
and the local value of $\gamma(0)$ and plot these on the [$t$-$\gamma(0)$]
plane of Fig. \ref{regionstgamma}.

More specifically, instead of performing this analysis for
absolute values of physical parameters, we first identify new dimensionless
variables for the temperature and interaction strength that are more
suitable under these conditions. The new parameters we introduce are the
global interaction parameter $\tilde{\gamma}$ and the global reduced
temperature $\theta$:
\begin{equation}
\tilde{\gamma}=\left( \frac{mg^{2}/(2\hbar^{2})}{N\hbar\omega_{z}}\right)
^{1/2},  \label{tildegamma-def}
\end{equation}
\begin{equation}
\theta=T/T_{Q},  \label{theta-def}
\end{equation}
where $T_{Q}=N\hbar\omega_{z}$.

\begin{figure}[ptb]
\includegraphics[width=6cm]{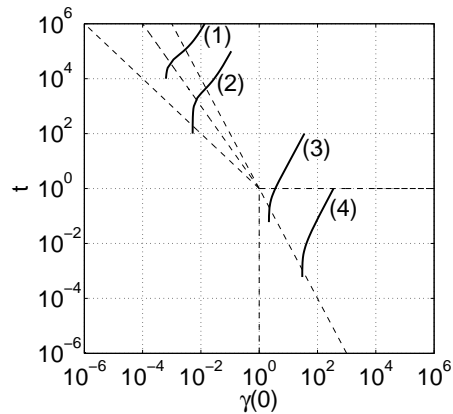}
\caption{Diagram of the regimes of a trapped $1D$ Bose gas as in Fig.
\protect\ref{regionstgamma}, except that the curved lines $(1)-(4)$
represent the locations of the interaction parameter $\protect\gamma(0)$ and
the reduced temperature $t$, for four different (fixed) values of the global
interaction parameter $\tilde{\protect\gamma}$ while the global temperature $
\protect\theta$ is changing. This represents four different samples with
fixed total number of particles $N$ and varying absolute temperature $T$.
For each point on a given line, there exists an associated density profile
with the peak density $n(0)$ corresponding to the respective value of $
\protect\gamma(0)$, and the local values $n(z)$ corresponding to the values
of $\protect\gamma(z)$ in the horizontal direction to the right. The values
of the global interaction parameter $\tilde{\protect\gamma}$ for each line
are: (1) -- $\tilde{\protect\gamma}=0.002$, (2) -- $\tilde{\protect\gamma}
=0.01$, (3) -- $\tilde{\protect\gamma}=1$, and (4) -- $\tilde{\protect\gamma
}=10$.}
\label{linesofconstN}
\end{figure}

The definition of the global interaction parameter $\tilde{\gamma}$, Eq. (
\ref{tildegamma-def}), relies on the identity
\begin{equation}
\frac{\theta}{\tilde{\gamma}^{2}}=\frac{\tau(0)}{\gamma^{2}(0)}=t.
\label{global-local-relationship}
\end{equation}
Using the definitions of the local parameters $\tau(0)$ and $\gamma(0)$,
 we see that $\tilde{\gamma}$ is the square root
of the ratio of two energy scales, $mg^{2}/(2\hbar^{2})$ and $T_{Q}=N\hbar
\omega_{z}$, as in Eq. (\ref{tildegamma-def}) (see also Ref. \cite
{tildegamma-comment}).

The definitions of the dimensionless temperature and
interaction parameters $\theta$ and $\tilde{\gamma}$ both include the total
number of particles $N$. This is more suitable for 
analyzing  the properties of the gas under conditions of changing
temperature at constant $N$.
In Fig. \ref{linesofconstN} we present the results of calculation of the
density profiles for four different (fixed) values of the global interaction
parameter $\tilde{\gamma}$ while the temperature $\theta$ is changed within
a broad range of values, typically between $0.1\lesssim\theta\lesssim10$.
The results are summarized by plotting the path of the resulting local
values of the interaction parameter $\gamma(0)$ at the trap center and the
reduced temperature $t$, in the parameter space of Fig. \ref{regionstgamma}.

\begin{figure}[ptb]
\includegraphics[width=6cm]{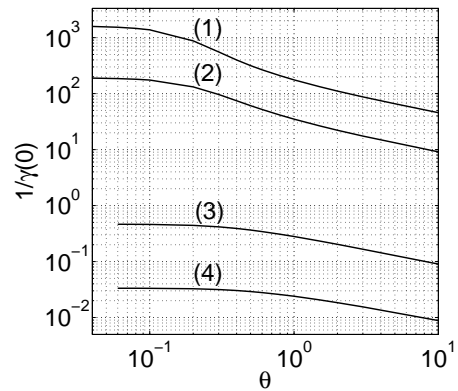}
\caption{Variation of the peak density $n(0)\propto 1/\protect\gamma(0)$ as
a function of the temperature $\protect\theta=T/N\hbar\protect\omega_{z}$ at
constant total number of particles $N$. This is the same data as in Fig.
\protect\ref{linesofconstN} except plotted in the $1/\protect\gamma(0)$-$
\protect\theta$ plane, where $\protect\theta=\tilde{\protect\gamma}^{2}t$.}
\label{theta-gamma0}
\end{figure}

For quantitative purposes, we also present the same data in the $\theta $-$
1/\gamma(0)$ plane, which is shown in Fig. \ref{theta-gamma0}. While Fig.
\ref{linesofconstN} identifies the local regimes of the gas with constant $N$,
Fig. \ref{theta-gamma0} helps to understand the properties of the gas in
terms of the variation in the global temperature parameter $\theta$. Note
that when $\tilde{\gamma}$ is kept constant, the variations in the
temperature parameters $\theta$ and $t$ are essentially equivalent and scale
as $\theta=$ $\tilde{\gamma}^{2}t$, according to Eq. (\ref
{global-local-relationship}).

There are simple approximate relations between the global and local
interaction parameters $\tilde{\gamma}$ and $\gamma(0)$ at high and low
temperatures. At high temperatures ($\theta\gg1$) the relationship is given
by:
\begin{equation}
\tilde{\gamma}\simeq\frac{\gamma(0)}{\sqrt{4\pi\theta}} ,\;\;\;\;\;\;\;\;\;
\;[\theta\gg1].  \label{tildegamma-gamma0-highT}
\end{equation}

At low temperatures ($\theta\ll1$), and in the limiting GP [$\gamma(0)\ll1$]
and TG [$\gamma(0)\gg1$] regimes, the relationship between $\tilde{\gamma}$
and $\gamma(0)$ becomes independent of $\theta$ and is given, respectively,
by:
\begin{align}
\tilde{\gamma} & \simeq\left( \frac{3}{8\sqrt{2}}\right) ^{1/2}
\gamma(0)^{3/4},\;\;\;\;\;[\theta\ll1,\;\gamma(0)\ll1], \\
\tilde{\gamma} & \simeq\frac{1}{\pi}\gamma
(0),\;\;\;\;\;\;\;\;\;\;\;\;\;\;\;\;\;\;\;\;\;\;\;[\theta\ll1,\;\gamma
(0)\gg1].
\end{align}

This implies that the GP and TG regimes can equivalently be defined via $
\tilde{\gamma}$ or $\gamma(0)$. The GP regime corresponds to $\tilde{\gamma }
\ll1$ or $\gamma(0)\ll1$, while the TG regime will correspond to $\tilde{
\gamma}\gg1$ or $\gamma(0)\gg1$.

From Fig. \ref{theta-gamma0} we see that at high temperatures and constant $
\tilde{\gamma}$, the local interaction parameter $\gamma(0)$ varies
according to $\gamma(0)\propto\theta^{1/2}$, i.e. linearly in the
logarithmic scale of Fig. \ref{theta-gamma0} and in agreement with Eq. (
\ref{tildegamma-gamma0-highT}). This means that the response of the peak
density $n(0)$ to temperature changes at constant $N$ follows the power law
of $n(0)\propto T^{-1/2}$, which is an expected result for the thermal
distribution of a classical ideal gas, Eq. (\ref{Thermal-Gaussian}).

As the temperature is reduced, the response of $n(0)$ to the temperature
changes becomes modified, and the modification is quite different depending
on the interaction strength $\tilde{\gamma}$. For weak interactions ($\tilde{
\gamma}\ll1$), as the temperature $T$ is reduced below $T_{Q}
=N\hbar\omega_{z}$ ($\theta=1$), the peak density $n(0)$ first increases
more rapidly than in a thermal gas, and then the growth is saturated as the
temperature is reduced further (see curves (1) and (2) in Fig. \ref
{theta-gamma0}). At very low temperatures ($\theta\ll1$), the peak density $
n(0)$ approaches a constant value independent of temperature. This is a
typical behaviour found in a weakly interacting gas that undergoes
quasi-condensation and reaches the GP regime. For intermediate and strong
interactions ($\tilde{\gamma}\gtrsim1$), on the other hand, the response of $
n(0)$ to the temperature reduction is different. Instead of an initial
speed up, the growth of the peak density $n(0)$ directly goes to the regime
of saturation, once the temperature is reduced below the global temperature
of quantum degeneracy $T_{Q}$ (see curves (3) and (4)). At very low
temperatures, $n(0)$ again approaches a constant value independent of the
temperature and the gas ends up in the TG regime.

From the paths of the curves (3) and (4) in Fig. \ref{linesofconstN}, we see
that achieving the TG regime from a high temperature classical gas by means
of reducing the temperature $T$ at constant $N$ requires large values of $
\tilde{\gamma}$ in the first place. This can be achieved by having a
relatively small total number of particles $N$ or a small trap frequency $
\omega_{z}$, according to Eq. (\ref{tildegamma-def}).

\section{Experimental considerations}

\subsection{Average pair correlation}

While the pair correlation $g^{(2)}(z,z)$ provides detailed information
about the local correlation properties of a trapped gas, its measurement as
a function of $z$ may not be an easy task in practice. Here, one usually
probes the pair correlation $\left\langle
\Psi^{\dagger}(z)\Psi^{\dagger}(z)\Psi (z)\Psi(z)\right\rangle$ within a
finite volume, e.g. via the measurement of the rates of two-body inelastic
processes within the entire sample. This means that one probes the
integrated or averaged correlation properties of the gas, as has been
demonstrated in a recent experiment of Ref. \cite{Tolra-NIST-exp}.

We are therefore motivated to study the average pair correlation defined via:

\begin{align}
\overline{G^{(2)}}& =\int dz\left\langle \hat{\Psi}^{\dagger }(z)\hat{\Psi}
^{\dagger }(z)\hat{\Psi}(z)\hat{\Psi}(z)\right\rangle  \notag \\
& =\int dzg^{(2)}(z,z)n^{2}(z).  \label{g2_av_def}
\end{align}

Here, an interesting question arises of whether this average correlation has
a simple relationship with the local pair correlation at the trap center $
g^{(2)}(0,0)$. The reason to expect this is the fact that $g^{(2)}(z,z)$ 
under the integral in Eq. (\ref{g2_av_def}) is
multiplied by $n^{2}(z)$ which vanishes rapidly as one approaches the tails
of the density profile. The function $g^{(2)}(z,z)$ near the trap center, on
the other hand, varies slowly and can be approximated by the value of $
g^{(2)}(0,0)$. Therefore, we can approximate $g^{(2)}(z,z)$ under the
integral by a constant $g^{(2)}(0,0)$, thus reducing Eq. (\ref{g2_av_def})
to
\begin{equation}
\overline{G^{(2)}}\simeq g^{(2)}(0,0)\int dzn^{2}(z).  \label{g2_av}
\end{equation}

\begin{figure}[tbp]
\includegraphics[width=6.5cm]{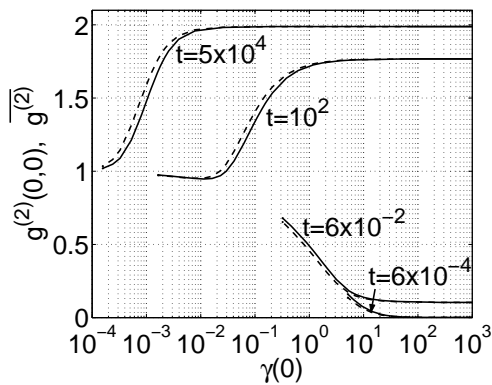}
\caption{The local pair correlation at the trap center $g^{(2)}(0,0)$ (solid
lines) and the normalized average pair correlation $\overline{g^{(2)}}$
(dashed lines) as a function of the interaction parameter $\protect\gamma
(0) $, for four different temperatures $t$. For each temperature, the
respective lines monitor the values of $g^{(2)}(0,0)$ and $\overline{g^{(2)}}
$ as one moves along the lines of constant $t$ in Fig. \protect\ref
{linesofconstt}.}
\label{g2_0}
\end{figure}

Thus, the average pair correlation $\overline{G^{(2)}}$ can be expressed via
the local pair correlation $g^{(2)}(0,0)$ using a simple relationship, 
Eq.~(\ref{g2_av}). Note that this also requires an independent evaluation of the
integral of the squared density, $\int dzn^{2}(z)$. Introducing a normalized
average pair correlation, $\overline{g^{(2)}}$, we obtain
\begin{equation}
\overline{g^{(2)}}\equiv \frac{\overline{G^{(2)}}}{\int dzn^{2}(z)}\simeq
g^{(2)}(0,0).
\end{equation}

In Fig. \ref{g2_0}, we plot the local pair correlation at the trap center $
g^{(2)}(0,0)$ and the normalized average pair correlation $\overline{g^{(2)}}
$ as a function of the interaction parameter $\gamma (0)$, for four
different temperatures $t$. Each line monitors the values of $g^{(2)}(0,0)$
and $\overline{g^{(2)}}$ as one moves along the lines of constant $t$ in
Fig. \ref{linesofconstt}. Here, the sequence of points along the horizontal
axis refers to the value of $\gamma (0)$ of the associated density profile $
n(z)$, for which we first calculate the pair correlation $g^{(2)}(z,z)$ as a
function of $z$ [which includes the plotted values of $g^{(2)}(0,0)$] and
then evaluate the integral in Eq. (\ref{g2_av_def}) to obtain the average
correlation $\overline{G^{(2)}}$, and hence $\overline{g^{(2)}}$. As we see,
in the limit of small $\gamma (0)$ the pair correlation approaches the
coherent level of fluctuations with $g^{(2)}(0,0)\simeq \overline{g^{(2)}}
\simeq 1$, while at large $\gamma (0)$ it can take any value between zero
and two, depending on the temperature $t$.

By comparing the full and dashed lines in Fig. \ref{g2_0}, we see that the
normalized average pair correlation $\overline{g^{(2)}}$ can indeed be well
approximated by the local pair correlation in the trap center $g^{(2)}(0,0)$. 
This is an important result and may have useful applications in practice.
For example, it gives a direct justification of the analysis performed in
Ref. \cite{Tolra-NIST-exp} where the results of the measurements of
three-body recombination rates in a bulk trapped sample have been compared
with theoretical predictions \cite{KK-DG-PD-GS-2003,DG-GS-NewJournalPhysics}
for a uniform gas.

\subsection{Practical example}

\begin{figure}[ptb]
\includegraphics[width=6cm]{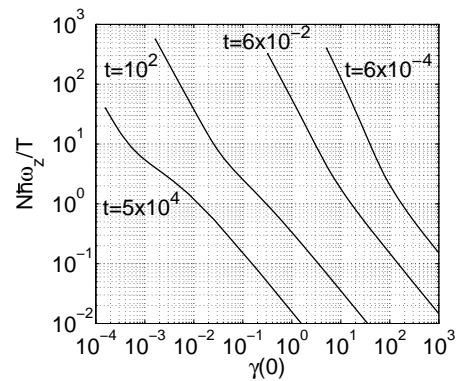}
\caption{Variation of $N\hbar\protect\omega_{z}/T$ as a function of $\protect
\gamma(0)$, for four different values of the temperature parameter $t$. For
a given coupling $g$ and a fixed $T$, this monitors the variation of $N\hbar
\protect\omega_{z}$ as one moves along the respective horizontal line in
Fig. \protect\ref{linesofconstt}. Here, each point along the horizontal line
is being referred to the value of $\protect\gamma(0)$ of the associated
density profile, for which we first calculate the density $n(z)$ as a
function of $z$ and then evaluate the resulting total number of particles $N=
\protect\int n(z)dz$ to form the dimensionless ratio $N\hbar\protect\omega
_{z}/T$.}
\label{N-gamma0}
\end{figure}

Here, we return to the analysis of Section V, with the reference to Fig. \ref
{linesofconstt}, and complete it by providing the results of calculation of
the total number of particles $N$. More specifically, we give the results
for the dimensionless ratio $N\hbar\omega_{z}/T$ as a function of the local
interaction parameter $\gamma(0)$ taken along the horizontal lines of Fig.
\ref{linesofconstt}, i.e. at four different (fixed) values of the
temperature parameter $t$. This is shown in Fig. \ref{N-gamma0}. For a given
trap frequency $\omega_{z}$ and coupling $g$, each line corresponds to
monitoring the variation in the total number of particles $N$ as a function
of the peak density $n(0)$, at constant temperature $T$.

Fig. \ref{N-gamma0} can also be viewed as giving the variation of the
interaction parameter at the trap center $\gamma(0)$ as a function of $
N\hbar\omega_{z}/T$, which in turn corresponds to monitoring the change in
the peak density of the gas $n(0)$ as the total number of particles $N$ is
varied at constant temperature $T$. Starting from the regime of low particle
numbers ($N\hbar\omega_{z}/T\ll1$ or high temperatures $\theta=T/N\hbar
\omega_{z}\gg1$), we see that the increase in $N$ results in a linear
increase of the peak density, $n(0)\propto N$. This is an expected result
for the thermal density distribution of a classical ideal gas, Eq. (\ref
{Thermal-Gaussian}), and corresponds to the linear dependence of $
N\hbar\omega _{z}/T$ on $\gamma(0)$ as seen in Fig. \ref{N-gamma0}.

As the number of particles is increased further and the ratio $N\hbar
\omega_{z}/T$ goes through the critical region $N\hbar\omega_{z}/T$ $\simeq1$
(corresponding to temperatures of the order of the global temperature of
quantum degeneracy, $T\simeq T_{Q}$) the growth of the peak density $n(0)$
speeds up, for the lines corresponding to $t=5\times10^{4}$ and $t=10^{2}$.
This speed up is most prominent in the first case corresponding to very week
interactions, and reflects the fact that the gas undergoes quasi-condensate
formation where the added particles mostly go into the trap center rather
than into the tails of the density distribution. With further increase of $
N\hbar\omega_{z}/T$, the growth of the peak density $n(0)$ is slower than in
the classical gas case and the situation is now reversed: the added
particles mostly go into the tails. Here, the gas is deep in the GP regime
and the density distribution is given by the inverted parabola, Eq. (\ref
{TF-GP-density}).

For the cases of $t=6\times10^{-2}$ and $t=6\times10^{-4}$, which at
constant $T$ correspond to stronger inter-particle interactions, the change
in the slop of the respective curves in Fig. \ref{N-gamma0} around $
N\hbar\omega_{z}/T$ $\simeq1$ represents the fact that the growth of the
peak density $n(0)$ only slows down when the number of particles is
increased past this critical region. Here, the gas goes first through the TG
regime where the density profile is given by the square root of inverted
parabola, Eq. (\ref{TF-TG-density}), and eventually it enters the GP regime
(see Fig. \ref{linesofconstt}) with gradual transformation of the shape of
the density profile to the GP parabola.

Apart from providing additional information about the properties of trapped
gases, Fig. \ref{N-gamma0} can also serve for quantitative analysis relevant
to practice. As an example, we consider a gas of $^{87}$Rb atoms ($
m=1.43\times10^{-25}$ kg, $a=5.3$ nm) with the aim of identifying a set of
physical parameters that correspond to the conditions of the point 3b in
Fig. \ref{linesofconstt}. Here, $\gamma(0)=2.16$ and $t=6\times10^{-2}$,
which is an example of a moderately ``fermionized'' gas. The pair
correlation $g^{(2)}(z,z)$ is below the coherent level $g^{(2)}(z,z)\ll1$,
yet the gas is not in the extreme low-temperature TG regime and therefore
the density profile can not be approximated by the square root of parabola.
In this sense, this example would be easier to realize in practice than the
extreme TG regime.

We first consider the trapping potential with $\omega_{z}/2\pi=20$ Hz and $
\omega_{\perp}/2\pi=80$ kHz ($l_{z}=2.42\times10^{-6}$ m and $
l_{\perp}=3.83\times10^{-8}$ m). With the $^{87}$Rb scattering length of $
a=5.3$ nm, giving $g\simeq2\hbar\omega_{\perp}a=5.62\times10^{-37}$ J/m, and
with the value of $\gamma(0)=2.16$ that we are aiming at, this set of
parameters results in the peak $1D$ density of $n(0)\simeq3.35\times10^{6}$ m
$^{-1}$ (and hence $n_{3D}\simeq1.82\times10^{20}$ m$^{-3}$). At this stage,
we can identify that the conditions of Eq. (\ref{1D-regime}) for achieving
the $1D$ regime are satisfied.

Next, the aimed value of $t=6\times10^{-2}$, together with the value of $g$
found above, gives the required absolute temperature $T\simeq1.22\times
10^{-31}$J (in energy units, or $T\simeq8.84$ nK). We next refer to the
results of Fig. \ref{N-gamma0} and read off the value of the dimensionless
ratio $N\hbar\omega_{z}/T\simeq16.7$ (or equivalently $\theta=T/N\hbar
\omega_{z}\simeq0.06$), which corresponds to $\gamma(0)=2.16$ on the
respective $t=6\times10^{-2}$ line. Finally, using the values of $\omega_{z}$
and $T$, we find that the required total number of particles here is $
N\simeq153$.

We note that all of the above parameter values are close to the
conditions realized in recent experiments \cite{Esslinger,Tolra-NIST-exp,
Weiss-exp,Bloch-and-Cazalilla}. 

\section{Summary}

In summary, we have obtained predictions for the correlations and density
profiles of a one-dimensional trapped Bose gas at finite temperature. This
allows previous results for the uniform 1D Bose gas to be applied to the
experimentally relevant case of a harmonic trap. The calculations use a
local density approximation which is asymptotically correct in the limit of
a large trap with a sufficiently slowly varying trap potential. We find
that, in this limit, there is a similar classification of different
coherence regimes as in the uniform case.

Remarkably, the density variation across the trap does not cause a dramatic
change in the average correlation function compared to the value at the trap
center. This is because the correlations are found to be relatively uniform
in the high density region near the trap center, which dominates any
trap-averaging measurement. This is particularly useful for experiments that
measure correlation functions through averaging a nonlinear interaction over
the length of the trap, which is the simplest currently available procedure.

We expect that direct measurements of $g^{(2)}$ that can test the
predictions of this fundamentally important many-body theory will become
feasible in the near future.

\begin{acknowledgments}
K.K. and P.D. acknowledge the Australian Research Council for the support of
this work. D.G. and G.S. acknowledge support from the Minist\`{e}re de la Recherche (grant ACI Nanoscience 201), from Centre National de la Recherche Scientifique (CNRS),
and from the Nederlandse Stichting voor Fundamenteel Onderzoek der Materie (FOM).
The research was also supported in part by the National Science Foundation under Grant No.
PHY99-07949. Laboratoire Kastler Brossel is a research unit  (UMR 8552) of
Universit\`{e} Pierre et Marie Curie and ENS, associated with CNRS. LPTMS is
a research unit (UMR 8626) of CNRS and Universit\'{e} Paris Sud.
\end{acknowledgments}


\begin{thebibliography}{99}
\bibitem{lc} Well below the temperature of quantum degeneracy $l_{c} =\hbar/
\sqrt{m\mu}$ in the GP and TG regimes ($\mu$ is the chemical potential), and
otherwise $l_{c}$ is of the order of $\Lambda_{T}$. In the GP regime $l_{c}$
coincides with the healing length, $l_{c}=\hbar/\sqrt{mng}$, and in the TG
regime $l_{c}\sim1/n$, where $n$ is the linear ($1D$) density.
\end{thebibliography}
\end{document}